\documentclass[1p,11pt]{elsarticle}

\usepackage{lineno,hyperref}
\modulolinenumbers[5]

\usepackage{amsmath}
\usepackage{amssymb}
\usepackage{newtxmath}
\usepackage{xspace} \usepackage{ifthen}
\usepackage{array}
\usepackage{xcolor}
\usepackage{url}
\usepackage{framed}
\usepackage{longtable}

\newboolean{figurelist}
\setboolean{figurelist}{false}

\newcommand	{\incfig}	[3]	{\ifthenelse{\boolean{figurelist}}
		{\immediate\write\outstream{fig-#1.pdf}}
	{}
\begin{figure}[!t]
    \includegraphics
				[#2]	{fig-#1}
    \caption{#3}
    \label{fig:#1}
	\end{figure}
	}

\renewcommand{\arraystretch}{1}
    
\newcommand {\doctable}
	[5] {
	\begin{table}[!t]
\scriptsize
	\caption{#2}
	\label{tbl:#1}
	\centering
	\begin{tabular}{#3}
#4
	\end{tabular}
	\end{table}
	}

\newenvironment
	{example}
	{
	\begin{description}
	\item [Example:]
	}
	{\end{description}}

\newenvironment{makefigurelist}
	{
\ifthenelse{\boolean{figurelist}}
		{
		\newwrite\outstream
		\immediate\openout\outstream=figure_list
}
		{} }
	{
\ifthenelse{\boolean{figurelist}}
		{
		\immediate\closeout\outstream
		}
		{}
	}

\newcommand{\eqnref}[1] {Eq.\eqref{eq:#1}}
\newcommand{\secref}[1] {\S\ref{sec:#1}}
\newcommand{\tblref}[1] {Tbl.\ref{tbl:#1}}
\newcommand{\figref}[1] {Fig.\ref{fig:#1}}

\newcommand{\ile}[1]{\mbox{$#1$}}

\newcommand{\sss}[2]{#1^{}_{\!_{#2}}}

\newcommand{\tH}{\sss{t}{H}}

\newcommand{\distanceTarget}{D}

\newcommand{\origin}{origin}

\newcommand{\destination}{destination}

\newcommand{\var}[1]{#1}

\newcommand{\trav}[1]{\var{\mathcal #1}}
\newcommand{\travtime}[1]{\var{\tau_{#1}}}

\newcommand{\coordtime}[1]{\var{t_{#1}}}
\newcommand{\coordistance}[1]{\var{x_{#1}}}

\newcommand{\direction}[2]{\trav{#1}{\Rightarrow}\trav{#2}}
\newcommand{\psidirection}[2]{\Psi_{\trav{#1}{\Rightarrow}\trav{#2}}}
\newcommand{\omegadirection}[2]{\Omega_{\trav{#1}{\Rightarrow}\trav{#2}}}

\newcommand{\latency}[2]{\var{\mathfrak L_{\trav{#1}{\Leftrightarrow}\trav{#2}}}}

\begin{document}

\begin{frontmatter}

\title{Timing relationships and resulting communications challenges in relativistic travel\tnoteref{t1}}
\tnotetext[t1]{Copyright\copyright 2023, all rights reserved.}

\author[1]{David Messerschmitt}
\address[1]{University of Calfornia at Berkeley,
Department of Electrical Engineering and Computer Sciences, USA}

\author[2,3]{Ian Morrison}
\address[2]{Curtin University, International Centre for Radio Astronomy Research, Australia}
\address[3]{Astro Signal Pty Ltd, Perth, Australia}

\author[4]{Thomas Mozdzen}
\address[4]{Arizona State University, 
Department of Physics, USA}

\author[5]{Philip Lubin}
\address[5]{University of California at Santa Barbara,
Department of Physics, USA}

\begin{abstract}
Communications between a spacecraft undertaking interstellar travel at near light speed to and from
its launch site (the origin) and a landing site (the destination), as well as with
other spacecraft, faces significant challenges.
Photon-based communication is significantly impacted by classical effects, including large photon propagation delay,
and relativistic effects, including time dilation experienced by clocks moving at high relative speeds.
The timing of communications by photon transfer,
as measured specifically by local clocks at origin and destination
and aboard spacecraft, is analyzed and
illustrated for concrete mission scenarios.
These include a spacecraft experiencing indefinite constant self-acceleration,
and a launch-landing mission, in which a spacecraft experiences constant acceleration
for the first half of its cruise phase and a like deceleration for the second half.
The origin and destination are assumed to be at rest
within a common inertial frame with a wide range of fixed distances separating them.
Several typical communication modes are considered, 
including one-way messaging, two-way message query with an expected response, 
and the one-way streaming of long program material such as a podcast or video.
The local-clock relative timing experienced by
the communicating entities including clock images (relation of transmit and receive clocks in one-way communication),
the query-response latency (the elapsed time between a query message and reception of a message in response),
and the time warping of a streaming program (nonlinear stretching or shrinking of the time axis) are included.
In particular, large query-response latency, except for
a short interval following launch or before landing, is a severe limit on remote control and social interaction.
When photons must travel in the same direction as the spacecraft,
communication blackouts strongly limit the periods of time during which communication is possible,
and restrict the opportunities for both one-way and two-way communication.
\end{abstract}


\end{frontmatter}

\onecolumn

\begin{makefigurelist}

\section{Introduction}

Consider a spacecraft cruising
between two at-rest locations (an \origin{} and a \destination{}) separated by large interstellar distances.
Maintaining communications
with the \origin{}, \destination{}, or with other spacecraft
brings many benefits, from practical mission engineering and management, to scientific data, 
and to the psychological benefits of astronauts remaining in touch with the civilization left behind. 
However, interstellar travel with human astronauts necessitates high spacecraft 
velocities approaching the speed of light. 
Relativistic effects thereby come into play, and these can have extreme 
(and sometimes surprising) impacts on communications systems and capabilities.

In particular, a central result of relativity is that clocks in relative motion run at different relative rates. 
This means that a relativistic traveler's measurements of time are inconsistent with
that of observers at the \origin{} or \destination{}. 
In isolation everyone is content 
with their own sense of time, but this affects the traveler's sense of connection to their environment.
This includes affecting communication 
among travelers in motion with relativistic relative velocities,
where clock inconsistency is directly apparent. 
In particular, it strongly affects perceptions of relative timing, 
which has to be taken into account 
in both the formulation and interpretation of communicated messages. 
This concept is explored analytically and quantitatively in this paper
and applied to two simple mission profiles.
See \secref{Nomenclature} 
for a compilation of variables employed
in the following.

\section{Travelers' clocks in relativistic travel}
\label{sec:localClocks}

Participants in interstellar travel, such as spacecraft (including instrumentation and possibly astronauts)
and civilizations that launch such spacecraft, 
perceive time in terms of clocks that they carry along with other payloads and supplies.
Suppose for one such participant $\trav{A}$ the time measured by a clock that it carries is $\travtime{a}$.
This is called the \emph{proper time} associated with $\trav{A}$, meaning ``the time
observed by a clock at rest relative to participant $\trav{A}$''.
Following \cite{RefNumber806} a more descriptive term for $\travtime{a}$ is $\trav{A}$'s \emph{traveler's time}.

\subsection{Clock images}
\label{sec:clockImage}

For many purposes 
there is no concern about other participants' traveler's times.
However any activity that involves two or more such participants, such as navigation
or communication of information among them,
emphasizes a relationship among their traveler's times.

Consider two spatially separated travelers in a one-dimensional space
with spatial dimension $x$.
These travelers are denoted by $\trav{A}$ and $\trav{B}$,
they have possibly distinctive velocities and accelerations as measured relative to $x$.

These travelers each carry their own clocks with them,
and these clocks measure traveler's times $\travtime{a}$ and $\travtime{b}$.
In a communication context (the primary concern here) 
this relationship between $\trav{A}$ and $\trav{B}$ is created by \emph{photon exchange},
in which $\trav{A}$ emits a photon at its traveler's time $\travtime{a}$ and that photon is
detected by $\trav{B}$ at its traveler's time $\travtime{b}$.
Thus $\trav{A}$ is in control, since this traveler chooses $\travtime{a}$,
and $\travtime{b}$ is determined by the
position of $\trav{A}$ at the instant of photon emission as well as the
time-dependent position of $\trav{B}$ (and the related measurement of its local clock) during the photon's propagation
through spacetime.

Define two related functions 
$\psidirection{A}{B}$ and $\omegadirection{A}{B}$
that establish a causal relationship between these times as
\begin{equation}
\label{eq:clockImage}
\travtime{b} = \psidirection{A}{B} \big[ \travtime{a} \big]
\text{ and } 
\travtime{a} =  \Psi^{-1}_{\trav{A} \Rightarrow \trav{B}} \big[ \travtime{b} \big]
= \omegadirection{A}{B} \big[ \travtime{b} \big]
\,.
\end{equation}
The interpretation of $\psidirection{A}{B}$ is the time $\travtime{b}$ at which $\trav{B}$ detects a photon
that was emitted by $\trav{A}$ at time $\travtime{a}$,
assuming that this photon propagates from $\trav{A}$ to $\trav{B}$,
assumed to be at the vacuum speed of light $c$.
The inverse $\omegadirection{A}{B}$ always exists because on physical grounds $\psidirection{A}{B}$ is
a strictly monotonically increasing function.
The domain and codomain (range) of $\psidirection{A}{B}$ is a subset of the real numbers.
The domain of $\omegadirection{A}{B}$ is the range of $\psidirection{A}{B}$ and vice versa.

The relationship \ile{\psidirection{A}{B}} is determined by kinematic variables
(relative positions, velocities, and accelerations of $\trav{A}$ and $\trav{B}$).
Similarly the inverse function 
\ile{\omegadirection{A}{B}=\Psi^{-1}_{\trav{A} \Rightarrow \trav{B}}} infers,
under the same conditions, the time of emission $\travtime{a}$
corresponding to an observed detection time $\travtime{b}$.
Function \ile{\psidirection{A}{B}} is called the \emph{clock image} \cite{RefNumber806}
because of its analogy to a picture or image of $\trav{A}$ taken by $\trav{B}$.
\begin{example}
If $\trav{A}$ wants to send $\trav{B}$ a ``happy birthday'' greeting, the relevant birthday is
defined in accordance with $\travtime{b}$.
If $\trav{A}$ wants that greeting (sent as a message carried by photons) to arrive at \ile{\travtime{b} = \travtime{r}},
corresponding to say 8:00am on $\trav{B}$'s next birthday, then $\trav{A}$
should originate that message at its own traveler's time \ile{\travtime{a} = \omegadirection{A}{B} \big[ \travtime{r} \big]}.
\end{example}

\subsection{Reversing direction of propagation}

The direction of the arrow in the designation $\direction{A}{B}$
specifies the direction of photon transfer from $\trav{A}$ to $\trav{B}$
and hence establishes a cause-and-effect relationship between the emission and detection of a
photon and the resulting timing relationship.
The alternative $\direction{B}{A}$ implies a transfer from $\trav{B}$ to $\trav{A}$
and is markedly different due to a
reversal of the direction of propagation for a photon relative to the motion of $\trav{A}$ and $\trav{B}$.
Thus $\psidirection{B}{A}$ cannot be inferred from  $\psidirection{A}{B}$
in any straightforward way; that is,
$\psidirection{A}{B}$ and $\psidirection{B}{A}$ must be determined independently.
\begin{example}
Continuing the last example, $\trav{B}$ receiving a birthday greeting message
at \ile{\travtime{b} = \travtime{r}} can infer that the message was
sent at $\trav{A}$'s traveler's time \ile{\travtime{a} = \omegadirection{A}{B} \big[ \travtime{r} \big] }.
Further if $\trav{B}$ immediately replies to that message at \ile{\travtime{b} = \travtime{r}},
then the time that message will arrive at $\trav{A}$, according to $\trav{A}$'s local clock,
can be inferred by $\trav{B}$ to be 
\ile{\travtime{a} = \psidirection{B}{A} \big[ \travtime{r} \big]}
(note the switch from \ile{\omegadirection{A}{B}} to \ile{\psidirection{B}{A}}).
Traveler $\trav{B}$ may want to take this arrival time into account in any return message,
such as relating it to $\trav{A}$'s own age at the traveler's time $\travtime{a}$ at which $\trav{A}$
receives the message.
\end{example}

\subsection{Compact notation for clock image compositions}
\label{sec:compact}

In the following it will often be the case that a given scenario of interest involves two or more photon
exchanges.
Anticipating that need, we employ a standard notation for the composition of functions.

\subsubsection{Communication among three travelers}
\label{sec:three}

This is illustrated by a configuration in which there are
three travelers \ile{\{\trav{A},\trav{B},\trav{C}\}} along the spatial axis $x$ in this order.
If $\trav{A}$ emits a photon at time $\travtime{a}$ in the $+x$ direction,
which is detected by $\trav{B}$ at time $\travtime{b}$, 
suppose that $\trav{B}$ in response to that detection immediately 
emits a photon which is in turn detected
by $\trav{C}$ at time $\travtime{c}$.
In that case the relationship among \ile{\{\travtime{a},\travtime{b},\travtime{c}\}} is summarized by
\begin{align}
&\travtime{b} = \psidirection{A}{B}[\travtime{a}]
\\
&\travtime{c} = \psidirection{B}{C}[\travtime{b}]
\\
\implies &\travtime{c} = \psidirection{B}{C}\big[\psidirection{A}{B}[\travtime{a}]\big]
\,.
\end{align}
In the following, we denote the composition of two clock images by the equivalent but arguably more
transparent representation
\begin{equation}
\travtime{c} = \big( \psidirection{B}{C} \circ \psidirection{A}{B} \big) [\travtime{a}] \,.
\end{equation}
With the composition operator ``$\circ$'', the invocation of the functions is from right to left.
Note that there is an assumption of immediacy; namely, $\trav{B}$ responds to its own detection of a photon
by immediately emitting a new photon.
In addition, the domain and range of the clock image functions has to be considered,
since it is possible to establish nonsensical conditions that violate these constraints. 
\begin{example}
$\trav{A}$ wants to wish $\trav{B}$ a happy birthday, with the message arriving at $\trav{B}$'s
traveler's time \ile{\travtime{b} = \travtime{r}} appropriate for the birthday.
As in the earlier example, $\trav{A}$ should originate that message at time 
\ile{\travtime{a} = \omegadirection{A}{B}[\travtime{r}]}.
Now suppose $\trav{A}$ wishes to remind a third traveler $\trav{C}$ that it is $\trav{B}$'s birthday
by sending to $\trav{C}$ a message ``Please immediately send a birthday greeting to $\trav{ B}$,
and if you do so it will arrive at the appropriate time''.
To satisfy that promise,
$\trav{C}$ should receive $\trav{A}$'s message at \ile{\travtime{c} = \omegadirection{C}{B} [\travtime{r}]}.
Thus $\trav{A}$ should originate that reminder message to $\trav{C}$ at its own traveler's time
\begin{equation}
\label{eq:threeTravBirthday}
\ile{\travtime{a} =
\omegadirection{A}{C}[\travtime{c}] =
\omegadirection{A}{C}\big[\omegadirection{C}{B} [\travtime{r}]\big] =
 \big( \omegadirection{A}{C} \circ \omegadirection{C}{B} \big) [\travtime{r}]} \,.
\end{equation}
This requirement becomes nonsensical if either $\travtime{r}$ falls outside the domain of function
$\omegadirection{C}{B}$, or if \ile{\travtime{c} = \omegadirection{C}{B} [\travtime{r}]} falls outside the
domain of function \ile{\omegadirection{A}{C}}.
This will occur whenever \ile{\{\trav{A},\trav{B},\trav{C}\}} are too distant
and the traveler's clocks are such that it is impossible
to get one or more messages to the appointed destinations by the appointed time.
\end{example}
The ``$\circ$'' version in \eqnref{threeTravBirthday} is quicker and easier to interpret,
and with a bit of practice also becomes straightforward to specify directly.

\subsubsection{Query-response communication}
\label{sec:query}

Another common situation is \emph{query-response} communication between two travelers.
Suppose that $\trav{A}$ wishes to receive a response from $\trav{B}$ to its own query.
This involves two one-way communications, first $\direction{A}{B}$ conveying the query
followed by $\direction{B}{A}$ conveying the response back to $\trav{A}$.
Suppose the original query traveler's time is at traveler's time $\travtime{a1}$, $\trav{B}$ 
responds immediately, and the response arrives back at $\trav{A}$ at
traveler's time $\travtime{a2}$.
Then we can infer that
\ile{\travtime{a2} = \big( \psidirection{B}{A} \circ \psidirection{A}{B} \big) [\travtime{a1}]}.
Note that the direction of photon travel is reversed in the two invocations of clock image $\Psi$.

Typically $\trav{A}$ will be interested in the elapsed time \ile{\big( \travtime{a2}-\travtime{a1} \big)}
(as measured by its local clock)
between the time it initiates a query and the time that it receives a response.
This suggests a new function called the \emph{query-response latency}, 
which is denoted by $\latency{A}{B}[\tau]$,
\begin{align}
\notag
\latency{A}{B}[\travtime{a1}] &=  
\big( \psidirection{B}{A} \circ \psidirection{A}{B} \big) [\travtime{a1}] - \travtime{a1}
\\
\label{eq:defnlatency}
&= \big( \psidirection{B}{A} \circ \psidirection{A}{B} - \vmathbb 1 \big) [\travtime{a1}] 
\end{align}
where $\vmathbb 1$ is the identify function.
This latency depends on the traveler's time $\travtime{a1}$ corresponding to the initiating query.
The notation \ile{\trav{A}{\Leftrightarrow}\trav{B}} denotes
a bi-directional communication between $\trav{A}$ and $\trav{B}$ that is initiated by $\trav{A}$.
\begin{example}
If $\trav{A}$'s query is ``how do I repair or circumvent my broken $\text{CO}_2$ scrubber'' then
circumstances favor a quick (low latency) response!
\end{example}
There is an implicit traveler's time \ile{\travtime{b} = \psidirection{A}{B}[\travtime{a1}]}
at which $\trav{B}$ receives the
query and is expected to immediately generate a response.
This value of $\travtime{b}$ will often influence $\trav{B}$'s response.
\begin{example}
If $\trav{A}$'s query is ``what time do you have'', the response
would be something like ``my current time is \ile{\travtime{b}= 1.645\text{ yr}} following my launch''.
From that response, if $\trav{A}$ is knowledgeable of the navigation plan for $\trav{B}$ then
trajectory coordinates \ile{\{x_b,\coordtime{b}\}} of $\trav{B}$ at the instant of response could also be inferred
(see \secref{trajectory}).
\end{example}

\subsection{Factors determining clock images}

There are three factors that collectively influence and determine the clock image
\ile{\psidirection{A}{B}} in the context of communication by photon exchange:
\begin{description}
\item[\textbf{Propagation delay.}]
In photon exchange there is a light-speed propagation delay between emission and detection.
If $\trav{A}$ is the photon emitter, the delay depends on $\trav{A}$'s position
at the time of emission, but not on $\trav{A}$'s subsequent motion.
The delay then depends on $\trav{B}$'s motion throughout the photon propagation, since that motion
affects $\trav{B}$'s distance at the subsequent photon detection.
\item[\textbf{Relativistic time dilation.}]
In the event of any relative motion,
$\travtime{a}$ and $\travtime{b}$ advance at inconsistent rates.
\item[\textbf{Clock initialization.}]
 Should the clocks $\travtime{a}$ and $\travtime{b}$ have arbitrary initializations, this would result in an unknown or unspecified mutual offset.
To eliminate this ambiguity we assume that the clocks are \emph{synchronized}
in a sense that depends on
the mission context (see \secref{clockSynch}).
\end{description}

\section{Spacecraft trajectories}

In order to infer the clock image functions, it is necessary to know
the \emph{trajectory} of the two travelers exchanging photons.
In the case of a spacecraft traveler, 
its \emph{mission} describes this trajectory in the context of a concrete objective
(such as scientific investigation or exploration).
The clock image $\psidirection{A}{B}$ 
and query-response latency $\latency{A}{B}$ are powerful tools in understanding
the limits to communication in the context of an interstellar mission, one that has the objective
of reaching the vicinity of stars other than our Sun.

Communication challenges are best illustrated in the context of concrete interstellar mission examples.
The goal here is to emphasize communications to and from a spacecraft
traveling at relativistic speed (a significant fraction of the speed of light),
while avoiding complications wrought by complex mission scenarios.
Therefore
two simple (and yet recognizably idealized) trajectories and associated missions are considered in the following:
\begin{description}
\item[Indefinite acceleration.]
Following launch from some stationary \origin{}, the spacecraft and its payload experience a constant acceleration that continues indefinitely.
In this case ``acceleration'' is that measured by an accelerometer carried by the spacecraft.
\item[Canonical mission.]
The spacecraft is launched from a stationary \origin{} and subsequently lands on a stationary \destination{}.
To this end it experiences a constant acceleration for half the distance from \origin{} to \destination{},
followed by a constant deceleration (with the same magnitude) for the remaining distance.
\end{description}
The assumptions and details of these two trajectories are now considered further.

\subsection{Indefinite acceleration assumptions}

The \emph{self-acceleration} (sometimes called \emph{proper acceleration}) 
of a spacecraft is that measured by an accelerometer carried
by the spacecraft.
Significantly this is the acceleration imparted by the propulsion subsystem,
and as well is the acceleration experienced by the payload, including possibly a human crew.

\emph{Indefinite} acceleration makes the assumption of a constant
self-acceleration with a numerical value $\alpha$
starting at the instant of launch (from a state of rest) and continuing thereafter, with no defined ending time.
Of course maintaining acceleration indefinitely is not physically realizable
(since it would require infinite energy expenditure)
but this assumption results in a simple and analytically tractable \emph{hyperbolic} trajectory (see \secref{analysisRelativity}).
Further, a canonical mission trajectory is readily inferred based on
a limited-time initial segment of a hyperbolic trajectory combined with the symmetry of the acceleration profile
(see \secref{trajectoryCanonical}).

\subsection{Canonical mission architecture}
\label{sec:missionCanonical}

A typical interstellar mission has the objective of landing on some alien body, and is also
physically realizable since its total energy expenditure is finite.
In such a mission,
a spacecraft is presumed to be launched from one stationary body, the 
\emph{origin}, and to subsequently land on another stationary body, its \emph{destination}.
Typical origins and destinations are planets
(for example Earth or Moon for an \origin{} and an exoplanet for a \destination{}).

The \emph{canonical} mission is the simplest mission with this objective,
and is illustrated in \figref{launchLandingMission}.
A spacecraft is propelled during a \emph{cruise} phase
with a constant self-acceleration for the first half of the journey and a constant (equal-in-magnitude) deceleration 
for the second half of the journey,

Although the \origin{} and \destination{} will generally be in relative motion,
their relative velocity will generally be insignificant compared
to the speed of light.
Thus in our canonical mission scenario any relative motion between \origin{} and \destination{} is neglected and they are
assumed to be inertial (not accelerating) relative to a common inertial frame $S$.

Our primary concern here is communication between the spacecraft and either
\origin{} or \destination{} during the cruise phase.
Although such a mission is presently beyond human technological capabilities 
and available energy resources, it is still worthwhile and interesting to consider
(as we do here) the implications of timing relationships among these entities during the mission.
The physical obstacles to communications uncovered is one among many important considerations in determining whether
the large investments required to enable and execute interstellar travel are justified. 

In the case of a mission carrying human astronauts, those astronauts (as well as any
biological material they carry, such as plants, animals, or food) have to survive the 
accelerations endemic to the journey.
While equipment can be designed to withstand large accelerations or weightlessness,
biological material originating from a planetary environment is vulnerable
to conditions not endemic to that environment, including large accelerations
and weightlessness.
To address this,
in numerical examples we assume a so-called 1-$g$ mission, which chooses
a constant acceleration magnitude equal to Earth gravity, or \ile{\alpha = \pm g},
thereby creating a spacecraft environment that replicates the Earth's surface in this respect.

\incfig
	{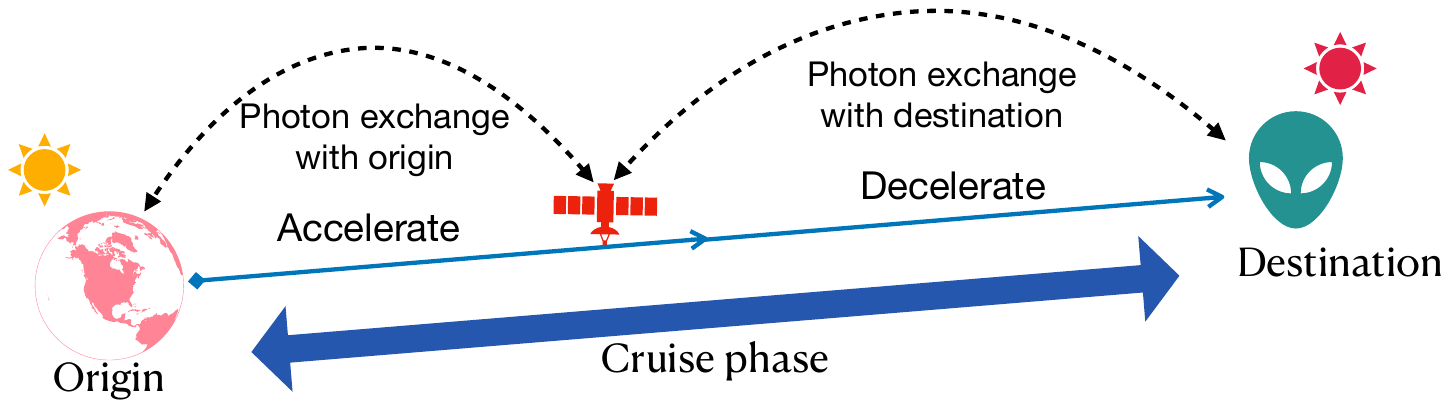}
	{
	trim=50 100 0 200,
    	clip,
    	width=.8\linewidth
	}
	{
     Illustration of the cruise phase of a canonical mission for a
	self-propelled spacecraft.
At rest at the \origin{}, beginning with launch it experiences
	constant self-acceleration for the first half of the distance to the \destination{}
	followed by a constant self-deceleration (with the same magnitude) for the second half,
	at rest at the instant of landing on the \destination{}.
	The \origin{} and \destination{} are themselves assumed to 
	be indefinitely at rest relative to a common inertial frame $S$.
}

\section{Relativistic effects on spacecraft trajectories}

When traveling near the speed of light $c$, relativistic effects strongly affect
the relationship between the spacecraft and its environment.

\subsection{Travelers' clocks}

Following the description in \secref{localClocks},
the canonical mission involves three travelers of interest:
the \origin{}  $\trav{O}$, the spacecraft $\trav{C}$, and the \destination{} $\trav{D}$.
(Although
$\trav{O}$ and $\trav{D}$ are assumed to be inertial,
they can still be considered ``travelers'' in a degenerate sense.)
The three travelers carry clocks measuring, respectively,
 traveler's times $\travtime{o}$, $\travtime{c}$, and $\travtime{d}$.
 (If this is not physically true of $\trav{D}$ because it is technology-sterile,
 this remains useful for analytical purposes).

\subsection{Inertial frame and coordinate times}

The location of the clocks of interest and a notation for their respective times are illustrated in
\figref{trajectory}.
The inertial frame $S$ defined previously
can be presumed to have 1-D spatial dimension $x$, as well as
coordinate time $t$ defined as the time measured by
any clock at rest with respect to $S$.
The speed of light as measured by \ile{\{x,t\}} is a constant $c$, and this
allows for the determination of propagation delays.
In the case of $\trav{O}$ and $\trav{D}$ their traveler's clocks are also at rest with respect to $S$,
and thus they can be considered to have coordinate times
\ile{\coordtime{o} \equiv \travtime{o}} and \ile{\coordtime{c} \equiv \travtime{c}}.
The coordinate time $t_x$ at an arbitrary intermediate position \ile{0 < x < \distanceTarget}
is defined as the time that would be measured by a clock at rest at that location,
and generally \ile{\travtime{c} \ne \coordtime{c}} due to the relative motion of the respective clocks.
All clocks of interest can have a fixed offset dependent on their initialization.
Taken together, the location of a traveler is quantified by its three coordinates
\ile{\{\tau,x,t\}}.

\incfig
	{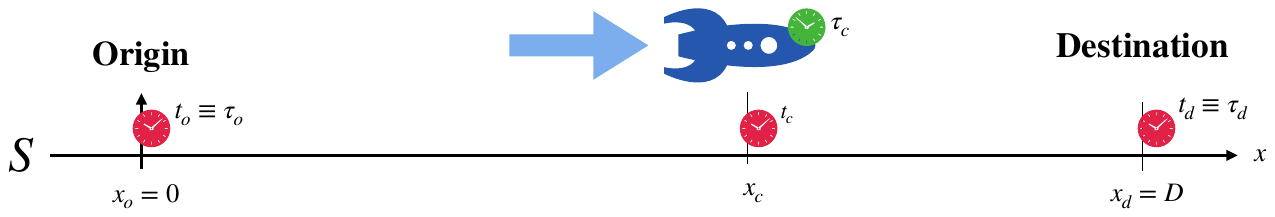}
	{
	trim=40 370 100 130,
   	clip,
    	width=1\linewidth
	}
	{
Illustration of the coordinates of a spacecraft's trajectory as a function
	of distance $x$ and coordinate time $t$ with respect to
	inertial frame $S$.
	The spatial coordinate $x$ of the \origin{}, spacecraft,
	and \destination{} are labeled
	 \ile{ \{x_o = 0, \coordistance{c}, x_d = \distanceTarget \}}.
	Clocks which measure coordinate times $t$ 
	at those positions measure times labeled
	 \ile{\{\coordtime{o},\coordtime{c},\coordtime{d}\}}.
	While  $\coordtime{o}$ will likely be
	measured by a physical clock, 
	\ile{\{\coordtime{c},\coordtime{d}\}} may be notional (rather than physical) clocks.
	In addition there are three traveler's (proper) times 
	\ile{\{\travtime{o},\travtime{c},\travtime{d}\}} carried by the respective travelers.
	The coordinate and travelers' clocks can be considered to be one and the same
	at the \origin{} and \destination{},
	but differ for the spacecraft position $\coordistance{c}$ due to their relative motion.
}
	
\subsection{Clock synchronization}
\label{sec:clockSynch}

While \ile{\{ \coordtime{o},\coordtime{c},\coordtime{d} \}} run at identical rates 
(since they possess no relative motion), 
there remains the question of their appropriate initialization,
and likewise the initialization of traveler's times \ile{\{\travtime{o},\travtime{c},\travtime{d}\}}.
The presumption of \ile{\travtime{o} \equiv \coordtime{o}} 
and \ile{\travtime{d} \equiv \coordtime{d}}
defines the synchronization of two of the six clocks of interest.
It is natural to assume that \ile{\coordtime{o} = \travtime{o} = \travtime{c}{=}0} at the instant of launch,
since the respective clocks are co-located and thus can be simultaneously initialized.

The remaining two times \ile{\{\coordtime{c},\coordtime{d}\}}
are not measured by clocks co-located with one another or with $\coordtime{o}$,
and thus their initialization is more challenging.
We assume that both are \emph{synchronized} with $\coordtime{o}$,
meaning that a measurement of the light-speed $c$ using
\ile{\{ \coordtime{o},\coordtime{c},\coordtime{d} \}} yields the correct value $c$.
This synchronization is also suitable for characterizing clock images.
That is, a photon emitted at time \ile{\coordtime{o} = 0} at the \origin{}
is detected at time 
\ile{\coordtime{d} = \coordistance{c} /c} at $\coordistance{c}$ and at
\ile{\coordtime{d} = \distanceTarget/c} at the \destination{}.
\begin{example}
If $\trav{O}$ sends a message to $\trav{D}$ as it initializes its local
clock to \ile{\coordtime{o} = 0}, that message traveling at the speed of light can upon arrival trigger an initialization
of the \destination{} clock to \ile{\coordtime{d} = \distanceTarget/c},
assuming that $\trav{D}$ has knowledge of message propagation distance
$\distanceTarget$.
\end{example}
When the propagation distance is unknown to the travelers, clock synchronization
requires bi-directional photon exchange protocols \cite{RefNumber806}.

\subsection{Measures of mission elapsed time}
\label{sec:MET}

The \emph{mission-elapsed time} (MET) is the cumulative time elapsed for the spacecraft
to reach some position $x_c$ during the spacecraft cruise phase.
There are two versions, depending on what clock is used to measure that elapsed time.
The mission elapsed coordinate time (MECT) is measured by coordinate time $\coordtime{c}$, 
whereas the mission-elapsed traveler's time (METT) is measured by the spacecraft's traveler's time $\travtime{c}$,
where we follow the convention that \ile{\travtime{c}{=}0} at the instant of launch.

The spacecraft speed must be less than light-speed $c$ relative to $S$, and thus
\ile{\text{MECT} > \coordistance{c}/c}.
Due to relativistic time dilation, the spacecraft's traveler's clock runs more slowly
than its respective coordinate clock, and thus \ile{\text{METT} \ll \text{MECT}} (see \secref{SRreview})
Thus it is entirely possible that \ile{\text{METT} \ll \coordistance{c}/c}, so that,
as measured by the traveler's clock, the light-speed limit has been exceeded.

An alternative trajectory  could include a ballistic
(constant speed) phase
in its center.
In this case the ballistic  phase would be bracketed by equal
durations of acceleration and deceleration.
Our canonical mission includes no such ballistic phase.
This alternative would have
minimal impact on the MECT at the time of landing, because
with or without the ballistic phase the spacecraft speed is very near to $c$ for most of the cruise.
However, as will be seen, omitting the ballistic phase beneficially and substantially reduces METT 
(to the benefit of resources consumed on the
spacecraft and biological aging).

\section{Communication with a spacecraft}
\label{sec:relativisitc}

Especially for a spacecraft with human astronauts,
persistent communication during the entire mission 
would be desirable during the cruise phase,
as opposed to following landing only.
However, except for relatively brief time periods shortly after launch and before landing,
the spacecraft will be traveling at nearly the speed of light.
This presents special
issues as we will see.  
Post-landing the spacecraft is assumed to be at rest in $S$,
so the situation is far simpler.

During the cruise phase there are four instances of communication represented by
\ile{\direction{O}{C}}, \ile{\direction{C}{O}}, \ile{\direction{C}{D}}, \text{and} \ile{\direction{D}{C}}.
For each of these cases the clock image $\Psi$ and its inverse $\Omega$
characterize the timing of the communication as affected by both propagation delay
and relativistic time dilation
(see \secref{clockImage}).

The query-response latency $\latency{A}{B}$ characterizes the delay experienced by a traveler
in a two-way communication.
Instances of these involving the spacecraft include
\ile{\trav{O}{\Leftrightarrow}\trav{C}},
\ile{\trav{C}{\Leftrightarrow}\trav{O}},
\ile{\trav{C}{\Leftrightarrow}\trav{D}},
and
\ile{\trav{D}{\Leftrightarrow}\trav{C}}.
This latency can be determined by the composition of one clock image $\Psi$ for the query
and another for the response.
Since travelers may not be fully cognizant of one another's trajectories,
both query and response messages are likely to
include a time-stamp confirming or revealing the local traveler's time.

Here we summarize the results of this analysis absent mathematical justification,
followed in \secref{analysisRelativity} by an analysis 
employing the special theory of relativity (SR).

\subsection{Spacecraft trajectory}
\label{sec:trajectory}

The spacecraft trajectory reveals the evolution of its position $\coordistance{c}$ in $S$ vs time,
specifically \ile{\{\travtime{c}, \coordistance{c},\coordtime{c}\}}.
This can be interpretated
as space time \ile{\{\coordistance{c},\coordtime{c}\}}
relative to $S$ coordinates as parameterized 
by the traveler's time $\travtime{c}$.
$S$ provides a common reference for measuring photon propagation delays,
time dilation effects for the spacecraft traveler's clock, and clock images.

In the case of a spacecraft, $\travtime{c}$  governs the operation of
the spacecraft instrumentation and is very significant to any human astronauts
(governing their heart rate, circadian rhythm, etc).
On the other hand, $\coordtime{c}$ matters greatly to navigation systems aboard the spacecraft
and at the \origin{}
(which measure and control spacecraft kinematics relative to an inertial frame $S$),
to controllers and operators at the \origin{} and \destination{}, and determines photon propagation delay.

\subsection{Event horizon}
\label{sec:eventHorizon}

The simpler indefinite self-acceleration trajectory is illustrated in \figref{trajConstAccel}.
A convenient parameter describing this trajectory is the \emph{event horizon}
\ile{\tH = c/\alpha}.
The physical significance of $\tH$ is the earliest coordinate time for which a
photon leaving the \origin{} never intersects the spacecraft trajectory,
and hence will never be detected onboard the spacecraft.

\incfig
	{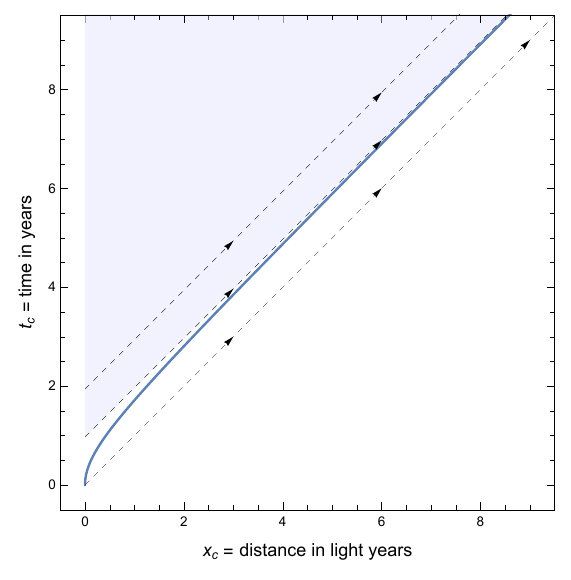}
	{
	trim=0 0 0 0,
    	clip,
    	width=0.6\linewidth
	}
	{
The trajectory of an indefinite 1-$g$ self-acceleration spacecraft
	with  \ile{\alpha_c = + g}
	and event horizon \ile{\tH = 0.97\text{ yr}}
	is plotted over an elapsed traveler's time of \ile{0 \le \travtime{c} \le 3\text{ yr}}.
	See \secref{trajectoryIndefinite} for a derivation of this trajectory.
	Also shown as dashed lines are the trajectories of a photon emission from the \emph{origin}
	at three coordinate times \ile{\coordtime{o} \in \{0,\tH,2 \tH \}} in the $+x$ direction.
	Each is a straight line with slope $+1$ on a plot with time units of years and distance units of light years.
}

Three photon trajectories (each linear with unity slope) are compared to the spacecraft in \figref{trajConstAccel}.
The first is a photon emitted from the \origin{} at the instant of spacecraft launch. 
The spacecraft has to travel more slowly than this photon, since it cannot exceed lightspeed $c$,
and thus its trajectory always falls above the photon's trajectory
(its coordinate time to reach any position $x_c$ has to be larger).
As the spacecraft speed approaches light-speed $c$, its trajectory exponentially approaches
(but never intersects)
that of a photon emitted from the \origin{} at \ile{\coordtime{o} = \tH}.
The trajectory of any photon emitted from the \origin{} for  \ile{\coordtime{o} \ge \tH},
which falls inside the shaded region on \figref{trajConstAccel},
never intersects the spacecraft trajectory, and hence no such photon can ever
be detected by the spacecraft.
A typical case is the photon emitted at  \ile{\coordtime{o} = 2 \tH} as shown.

In summary, photon emissions originating from the \origin{} beyond the event horizon
(shown as the shaded area)
are invisible to spacecraft experiencing indefinite acceleration.
The result is a communications \emph{blackout} from the \origin{} to the spacecraft.
While such a blackout is not present for the canonical mission,
the value of $\tH$ has a significant impact on clock images in that case as well.

\subsection{Spacecraft communication with \origin{} and \destination{}}
\label{sec:fourTraj}

The trajectory of a spacecraft on the canonical mission is illustrated in \figref{trajEnsemble}
together with some typical trajectories for photons exchanged to and from
the \origin{} $\trav{O}$ and \destination{} $\trav{D}$.
Due to the time-reversal
symmetry of the acceleration/deceleration,
the second half of the canonical mission trajectory is simply a mirror image of the first half
(see \secref{trajectoryCanonical}).

\begin{figure}[!t] 
	\begin{minipage}[t]{6cm} 
		\centering 
		\includegraphics[scale=0.7]{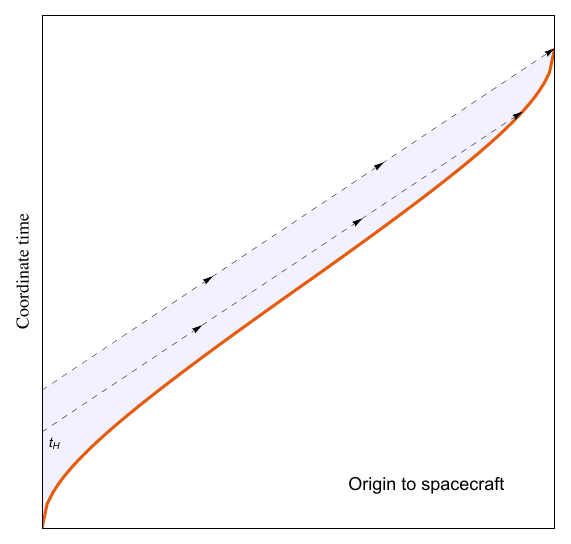} 
	\end{minipage} 
	\hspace{1cm}
	\begin{minipage}[t]{6cm} 
		\centering 
		\includegraphics[scale=0.7]{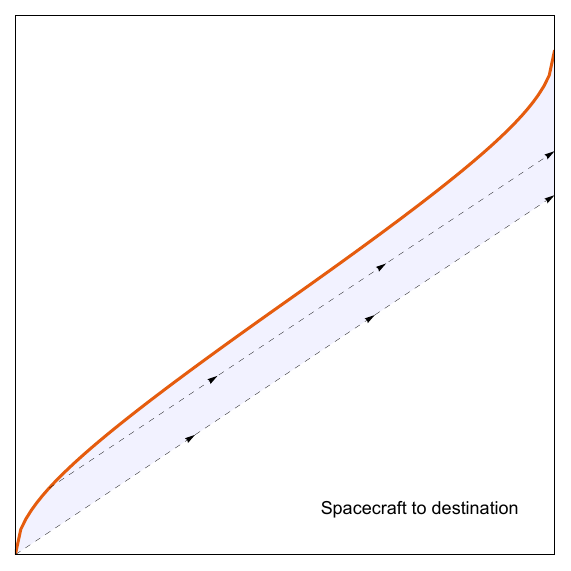} 
	\end{minipage} 
	\\[.2cm]
	\begin{minipage}[t]{6cm} 
		\centering 
		\includegraphics[scale=0.7]{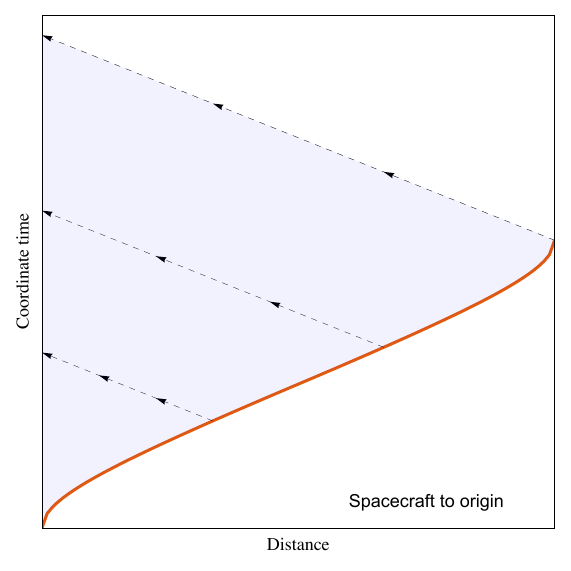} 
	\end{minipage} 
	\hspace{1cm}
	\begin{minipage}[t]{6cm} 
		\centering 
		\includegraphics[scale=0.7]{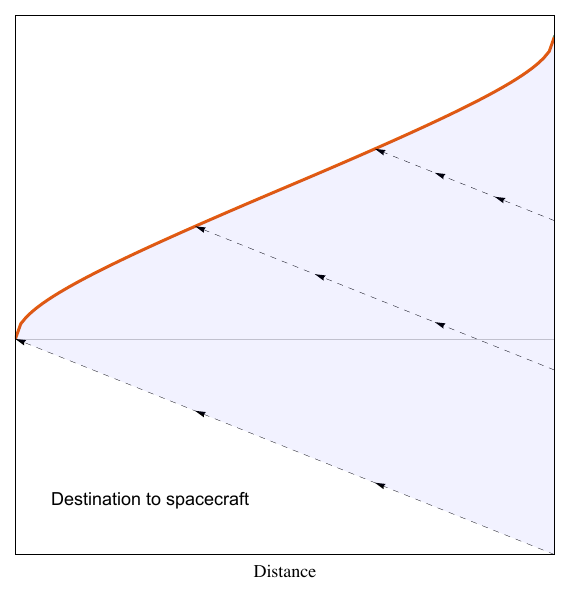} 
	\end{minipage} 
	\caption{
Illustration of the time constraints on communication during
	a spacecraft's travel from launch to landing for the canonical mission.
	The spacecraft trajectory is the solid line, and photon trajectories
	to and from the spacecraft and to and from the \origin{}
	and \destination{} are shown as dashed lines.
	Each shaded region represents the ensemble of all
	feasible photon trajectories which intersect the spacecraft trajectory during its cruise phase.
	The most significant influence on this ensemble shape is the direction of photon propagation. 
	}
	\label{fig:trajEnsemble}
\end{figure}

The shaded regions in \figref{trajEnsemble} represent photons that intercept
$\trav{C}$'s trajectory, and thus were either omitted or can be detected during $\trav{C}$'s cruise phase.
These photons are deemed to be \emph{productive},
in the sense that they can be employed as a basis for communication during $\trav{C}$'s cruise phase.
Photon trajectories outside these shaded regions must be emitted or
detected while $\trav{C}$ is at rest at $\trav{O}$  pre-launch or at $\trav{D}$ post-landing.
$\trav{C}$'s payload (including possibly human astronauts)
is governed by a local clock measuring traveler's time $\travtime{c}$
(represented only by parameterization in \figref{trajEnsemble}).

In the direction $\direction{O}{C}$,
a photon emitted from the \origin{} over the coordinate-time interval \ile{0{<}\coordtime{o}{< }{\sim}2 \tH}
can be detected by the spacecraft during its cruise phase
(later-emitted photons reach the spacecraft following landing at the \destination{}).
The factor ${\sim}2 \tH$ is the composition of
${\sim}\tH$ for the first half and another ${\sim}\tH$ (due to the mirror-image symmetry) for the second half.
(The precise range of times is slightly smaller than ${\sim}\tH$, since the positive self-acceleration of the spacecraft
is terminated at the halfway point.)

A set of important distinctions and conclusions can be drawn from \figref{trajEnsemble}:
\begin{itemize}
\item
From the perspective of spacecraft $\trav{C}$, detection of photons
originating from both \ile{\{\trav{O},\trav{D}\}} and emission of photons toward both \ile{\{\trav{O},\trav{D}\}}
are productive during the entirety of the cruise phase.
This constitutes $\trav{C}$'s entire cruise-phase METT.
\item
Consider the perspective of $\trav{O}$ and $\trav{D}$ and
photons either emitted or detected that propagate in the same direction
as $\trav{C}$'s velocity (which is the direction $\trav{O}$ to $\trav{D}$).
That includes directions $\direction{O}{C}$ and $\direction{C}{D}$.
These photons are productive
during a relatively short coordinate-time duration ${\sim}2 \tH$ in both cases.
Since \ile{\text{MECT} = {\sim}\big(\distanceTarget/c{+}2 \tH \big)} for the entirety of the cruise phase,
there is a long coordinate-time duration \ile{{\sim}\distanceTarget/c} 
overlapping the MECT over which no photon emissions
and detections can be productive.
\item
Consider again the perspective of  $\trav{O}$ and $\trav{D}$,
but in this case photons either emitted or detected propagate in the opposite direction
to the spacecraft $\trav{C}$'s velocity.
This includes directions $\direction{C}{O}$ and $\direction{D}{C}$.
Those photons
are productive during a long coordinate-time duration
which is about \emph{double} the METT.
\item
There is a substantial distinction between the
productive coordinate-time duration of communication in both directions
as compared to the spacecraft's METT.
This represents a \emph{time-warping} effect, in which communication signals always appear to have
distinctive time axes at transmitter and receiver (see \secref{streaming}).
Time-warping is distinct from relativistic time-dilation (see \secref{SRreview}), and is considerably more extreme
in its effect.
While time-warping is affected by time-dilation, it is primarily
a manifestation of the observation that photons propagating in the same direction as the spacecraft
have difficulty ``catching up with'' or ``getting out ahead of'' a spacecraft 
traveling at near the speed of light $c$.
\end{itemize}

The MECT at landing, from the perspective of the \origin{} $\trav{O}$ or \destination{} $\trav{D}$,
defines a coordinate-time period coinciding with the cruise phase of spacecraft $\trav{C}$.
Ideally communication to and from the spacecraft would be possible during the entirety of this MECT.
Whenever such communication is not possible
(resulting from non-productive
photon emissions or detections),
this constitutes a \emph{communication blackout}, which is summarized in \tblref{cases}.
The practical manifestation of these blackouts is that $\trav{C}$ (or $\trav{D}$) cannot be informed of any events
at $\trav{O}$ ( or $\trav{C}$) except for a short period of coordinate-time duration  ${\sim}2 \tH$
(about two years for a 1-$g$ canonical mission)
immediately following launch (or preceding landing).

 \doctable
	{cases}
    	{Communication blackout conditions}
   	 {l p{10cm}}
   	 {
\\ \hline \\ [-1ex]
$\direction{O}{C}$
&
Communications from $\trav{O}$
does not reach the spacecraft $\trav{C}$ during the cruise phase except for a coordinate-time duration \ile{{\sim}2 \tH}
immediately following launch.
Such communications do arrive following landing, while the spacecraft is at rest at the \destination{}.  
\\[15pt]
$\direction{C}{O}$ and $\direction{D}{C}$
&
Communication is possible during $\trav{C}$'s entire cruise phase.
\\[5pt]
$\direction{C}{D}$
&
$\trav{D}$ receives no communication from $\trav{C}$ initiated during its cruise phase
except for a coordinate-time duration ${\sim}2 \tH$ immediately preceding landing.
\\[15pt] \hline
      }

\subsection{Mission elapsed time at landing}
\label{sec:MECTT}

The mission elapsed times can be considered and compared from a traveler's and 
coordinate time perspective.
The traveler's perspective is simpler, 
since \ile{\text{METT}\equiv \travtime{c}} for cruising distance
\ile{0 < x \le \coordistance{c}}.
There are three coordinate times \ile{\{\coordtime{o},\coordtime{c},\coordtime{d}\}}
that differ in their initialization, 
but \ile{\text{MECT}\equiv \coordtime{c}} is the correct choice for the MECT
when the spacecraft is at position $x_c$,
since it defines the
coordinate time at that position.
The initializations of \ile{\{\coordtime{o},\coordtime{c}\}} are set
to correctly measure the speed of a photon, which implies that they
also correctly measure a photon's propagation time 
\ile{\coordtime{c}-\coordtime{o}} from \ile{x=0} to \ile{x=\coordistance{c}}.
It follows that those initializations are also the correct measure of
MECT for a spacecraft.

The MECT and METT at the time of landing are quite different.
The landing-time MECT of the canonical 1-$g$ mission is
\ile{\text{MECT} = {\sim}(\distanceTarget/c{+}2 \tH)}.
This is an asymptote for large $\distanceTarget$, with the precise value plotted in \figref{canonicalMECT}
((see \secref{fourTraj}).
The METT is plotted in \figref{canonicalMETT}, and is much smaller (\ile{\text{METT} \ll \text{MECT}})
due to relativistic time dilation.

\incfig
	{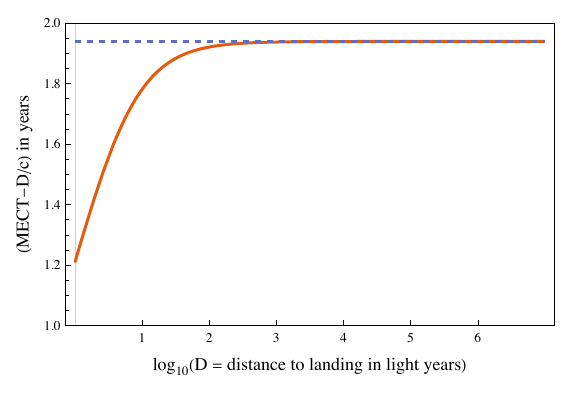}
	{
	trim=0 0 0 0,
    	clip,
    	width = 0.65\linewidth
	}
	{
A plot of
	the excess of the MECT over the time for a photon to reach the \destination{}
 for the canonical mission of \figref{launchLandingMission}.
	This is compared to twice the event horizon ($2 \tH$) as the dashed line.
	The horizontal axis is the log of distance $\log \distanceTarget$ 
	for \ile{1 \le \distanceTarget \le 10^7 \text{ ly}}.
	The longest distance exceeds the distance to the Andromeda galaxy.	
	At greater distances, twice the event horizon $2 \tH$
	(the dashed line) is a good approximation to that excess MECT.
}

\incfig
	{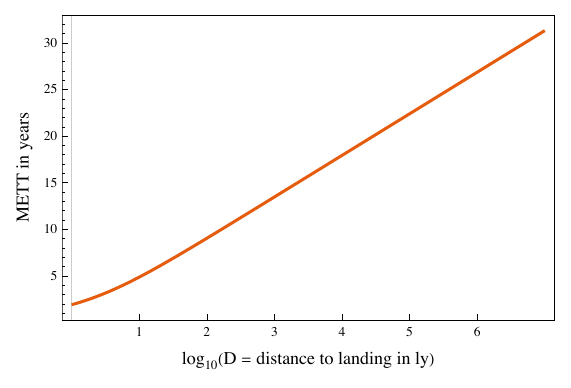}
	{
	trim=0 0 0 0,
    	clip,
    	width = 0.65\linewidth
	}
	{
	A repeat of \figref{canonicalMECT} for the traveler's time (METT).
	The canonical-mission METT falls well within a typical single human lifetime even
	at the greatest distance shown ($10^7$ ly).
	}

This distinction between MECT and METT is very significant for spacecraft designers, since METT
governs all time-dependent aging and resource utilization,
including the time period over which self-acceleration has to be maintained,
on-board energy,
wear and tear of equipment,
life support supplies for human astronauts including food and oxygen, 
and the biological aging of the astronauts.
Even for a journey to the nearest galaxy Andromeda 
(\ile{\distanceTarget{\approx}2.5 \cdot 10^6\text{ ly}})
the total canonical-mission METT at landing falls within a typical human lifetime.

Another consideration is the timekeeping aspect of communications.
This affects the perception of the communication from the receiving end,
but the transmitting end should also be aware of any time limitations on their
ability to convey information or events.
\begin{example}
Consider the uplink (direction $\direction{O}{C}$).
A spacecraft crew that travels \ile{\distanceTarget = 1000\text{ ly}} 
will have received, 
at the instant of landing at the \destination{},
\ile{{\sim}1.9\text{ yr}} 
of transmissions from the \origin{} (as measured by the \origin{}'s coordinate clock) during a \ile{{\sim}13.5 \text{ yr}} journey 
(as measured by the spacecraft's onboard traveler's clock).
Thus the crew perceives that they have lost visibility into \ile{{\sim}11.6 \text{ yrs}} of current
events at the \origin{} during their journey.
Thereafter the communication from the \origin{} will stream in real-time since both
transmitter and receiver are at rest in the same inertial frame $S$.
Thus the crew will, after landing, be kept up to date with events at the \origin{}
as they happen, although \ile{{\sim}11.6 \text{ yr}} after the fact
according to their own traveler's clock.
Any denizens at the \destination{} will, in contrast, have the perception that
their knowledge of events at the \origin{} is 1000 yr after the fact,
which is dramatically greater.
\end{example} 
The downlink (direction $\direction{C}{O}$) is very different due to the relatively rapid advance of the
coordinate-time.
Photons emitted
from the spacecraft will arrive at the \origin{} with an $\coordistance{c}/c$ propagation delay
as measured by the \origin{}'s coordinate clock,
which becomes large for large $\distanceTarget$.
\begin{example}
Continuing the last example, at the \origin{},
news of a successful landing will be delayed for \ile{{\sim}2002 \text{ yrs}}
following the spacecraft launch, which amounts to many generations for a human population.
During those intervening centuries,
\ile{{\sim}13.5 \text{ yrs}} of transmissions from the spacecraft will have been slowly received,
hopefully serving to maintain awareness of (and interest in) the mission and its outcome.
Thereafter, after landing the \origin{} will receive real-time transmissions from the spacecraft, albeit
with the perception that they are arriving \[{\sim}(2002-13.5) = 1988.5 \,\text{yr} \,\] after the fact.

Rather than communicate back to the \origin{},
an alternative scenario for the spacecraft would be to immediately turn around and travel back to the \origin{}.
This would effectively implement time travel \ile{{\sim}2004 \text{ yr}}  into the future,
at the expense of \ile{{\sim}27 \text{ yr}} total elapsed traveler's time.
\end{example}
A general conclusion of the last examples is that the impact of effects like propagation delay and time horizon
is minimal for the spacecraft payload in isolation, largely due to compensating relativistic time-dilation effects.
However, the impact of these conditions on denizens of the \origin{} and \destination{} is great, and as will be seen
(see \secref{queryResponse}), all parties encounter severe 
obstacles to two-way communication among themselves.

\subsection{One-way communication: Clock images}
\label{sec:imageDef}

The method of calculating the clock image function (see \secref{clockImage}) from two trajectories
is illustrated in \figref{illusClockImage}.
The trajectories are plotted with respect to inertial frame $S$, and photon trajectories
are plotted as unit-slope dashed lines.
Each photon trajectory that intersects the two traveler's trajectories
in the direction \ile{\direction{A}{B}} defines the clock image 
\ile{\travtime{b} = \psidirection{A}{B}(\travtime{a})}
for the pair of traveler's times \ile{\{\travtime{a},\travtime{b}\}} at the intersection with that photon trajectory.

\incfig
	{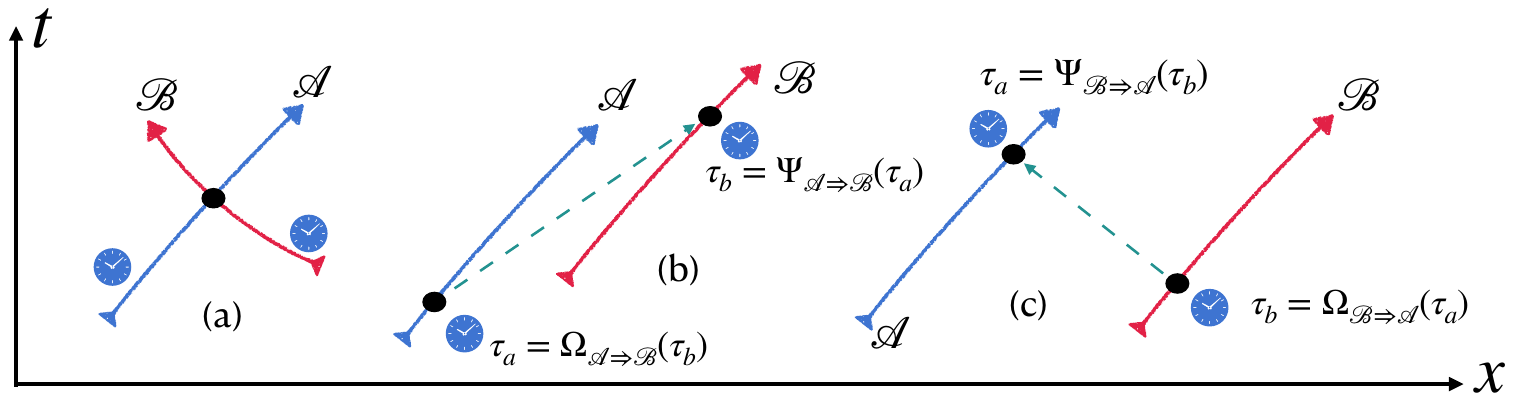}
	{
	trim=0 320 30 90,
    	clip,
    	width = 1\linewidth
	}
	{
	Illustration of the clock image functions for photon transfer between traveler $\trav{A}$
	with traveler's clock $\travtime{a}$ and traveler $\trav{B}$ with traveler's clock $\travtime{b}$
	with specified spacetime trajectories represented relative to a common inertial frame $S$.
    For the degenerate case of an \origin{} or \destination{} at rest, the corresponding trajectory will be vertical. 
	A photon trajectory in $S$ which intersects both trajectories creates a
	mapping  \ile{\travtime{a}{\Rightarrow}\travtime{b}} which can be represented by
	a clock image function $\psidirection{A}{B} (\cdot)$, as well as a mapping  
	\ile{\travtime{b}{\Rightarrow}\travtime{a}}
	which can be represented by an inverse function $\omegadirection{A}{B} (\cdot)$.
Note that one or both of the travelers may be at rest (say at the \origin{}
	or \destination{}), and for such a traveler the trajectory is vertical (with \ile{\tau \equiv t}
	within an arbitrary initialization).
(a) 
	If the trajectories intersect at some point in spacetime, the instantaneously 
	contiguous traveler's clocks can be directly compared, and if desired initialized to
	achieve synchronization.
	For communications this is an uninteresting degenerate case.
	(b)
	A photon trajectory in the $+x$ direction.
	(c)
	A photon trajectory in the $-x$ direction.
	}

Over the course of a mission there will be a considerable
time-warping
of the time base of the received signal transmissions in relation to the transmitter 
time axis (see \secref{streaming}).
This is manifested by the clock image functions shown in \figref{clockImage1gVsDistanceUp}
and \figref{clockImage1gVsDistanceDown} for a range of
launch-landing distances, and for an uplink from \origin{} to spacecraft
and downlink from spacecraft to \origin{}.

\incfig
	{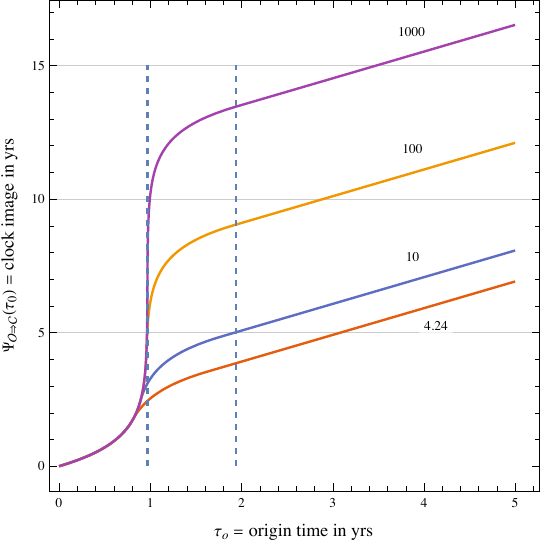}
	{
	trim=0 0 0 0,
    	clip,
    	width=.7\linewidth
	}
	{
A plot of the clock image \ile{\psidirection{O}{C} (\travtime{o})} in years 
	for photon transfer from \origin{} to spacecraft,
 for a canonical 1-$g$ mission. 
The vertical dashed lines mark the event horizon $\tH$ and twice the event horizon $2 \tH$,
	which are, respectively, upper bounds on the first half and entirety of the spacecraft cruise phase.
	Beyond ${\sim}2 \tH$ the clock image is linear with unit slope because the spacecraft is at rest at the \destination{}.
Both horizontal and vertical scales are linear (rather than logarithmic), 
	reflecting the relatively short METT experienced by the spacecraft
	as well as the relatively short interval before the event horizon is reached at the \origin{}
	(beyond that, transfer has to await the second half of the cruise phase or following landing).
	In practical terms, a reasonable delay is achieved for a relatively
	short time interval following launch.
	The different curves show the clock image function for different distances
	\ile{\distanceTarget \in \{4.24,10,100,1000 \} \text{ ly}} to the \destination{},
	with an expansion in METT for the longer distances.
}
	
\incfig
	{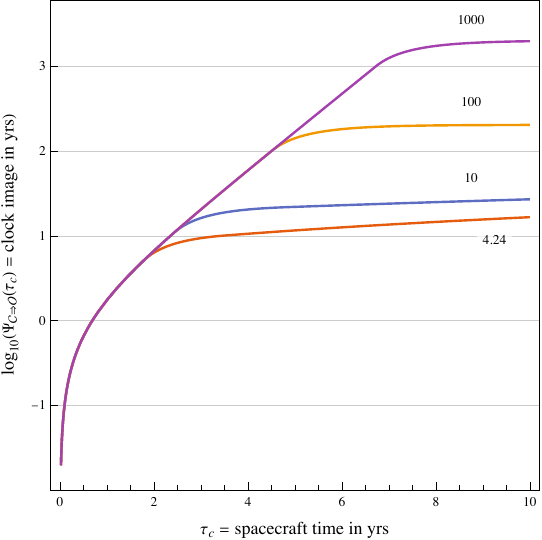}
	{
	trim=0 0 0 0,
    	clip,
    	width=.7\linewidth
	}
	{
The clock image \ile{\psidirection{C}{O}(\travtime{c})} for a photon transfer from
	a spacecraft to \origin{}.
	All the conventions of \figref{clockImage1gVsDistanceUp}  are repeated, except the
	direction of photon transfer is reversed.
	The image axis is now logarithmic (in contrast to \figref{clockImage1gVsDistanceUp})
	because photon transfers from the spacecraft now persist over a much longer period
	(reflecting \ile{\text{MECT} \gg \text{METT}}).
}

On the uplink $\direction{O}{C}$ shown in \figref{clockImage1gVsDistanceUp}, 
there is a sudden increase in the uplink clock image $\psidirection{O}{C}$ at the event horizon $\tH$
because photons will no longer intercept the spacecraft trajectory during the acceleration
phase and thus delay their arrival at the spacecraft to the deceleration phase.
Post-landing $\psidirection{O}{C}$ becomes linear with unit slope since the two
clocks increase in lockstep.
On the downlink $\direction{C}{O}$ shown in \figref{clockImage1gVsDistanceDown} this transition is more gradual, 
because the photon trajectory is oblique (rather than nearly tangent)
to the spacecraft trajectory, but $\psidirection{O}{C}$
is exponentially larger (requiring a log plot) since the 
\origin{} clock does not benefit from time dilation.
Thus the photon arrival latency (as observed by the local clock) is exponentially larger on the downlink.
As with the uplink, post-landing $\psidirection{C}{O}$
becomes linear.

\subsection{Two-way communication: Query-response latency}
\label{sec:queryResponse}

Much can be accomplished by one-way telemetry from spacecraft to \origin{}
(for example monitoring of the status of mechanical systems or human crew)
and a stream of current events from the \origin{} to the spacecraft.
However, if the spacecraft is not to be totally autonomous, 
there is need for two-way communication.
This can take the simple form of a query followed by a response to that query (see \secref{query}).
In the case of a `conversation', a query followed by a response followed by another query
following up on that response, these query-responses repeated indefinitely, might occur.
Such conversations are especially useful to human astronauts
in maintaining a social connection with family and friends.

\subsubsection{Indefinite constant self-acceleration mission}

When the self-acceleration continues indefinitely, there are only two
query-response latencies of interest, $\latency{O}{C}$ and $\latency{C}{O}$.
The values are determined in \secref{imageQuery} and illustrated in
\figref{latencyIndefiniteAcceleration}
for a 1-$g$ mission.
Because either the response or the query has to propagate in the same direction as the spacecraft motion,
in both cases the traveler's time of a query is limited by the time horizon
and latency increases dramatically as that horizon is approached.

\incfig
	{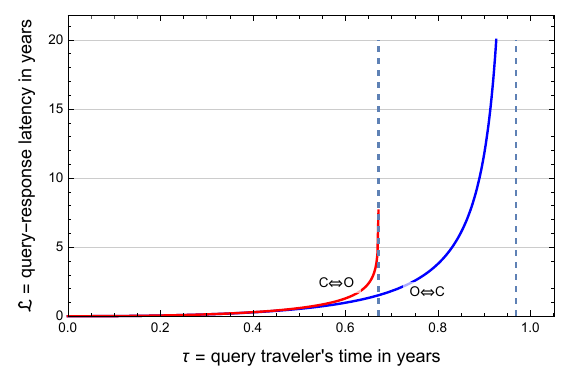}
	{
	trim=0 0 0 0,
    	clip,
    	width = 0.6\linewidth
	}
	{
The query-response latencies $\latency{O}{C}$ and $\latency{C}{O}$ 
	are compared
 for an indefinite self-acceleration mission with \ile{\alpha = g}.
	The opportunity for obtaining a response to a query originating at the
	\origin{} is limited to the time horizon \ile{\travtime{o} < \tH} (shown as  the second dashed line),
	and queries from the spacecraft are more severely limited to \ile{\travtime{c} < \tH \cdot \log 2}
	(also shown as the first dashed line).
	The latter restriction allows a query to reach the \origin{} soon enough that the
	response is generated prior to $\tH$.
}

\subsubsection{Canonical mission}

For the canonical mission, which adds the possibility of communicating with the \destination{},
there are six distinctive query-response latencies,
\[\{\latency{O}{C},\latency{C}{O},\latency{C}{D},\latency{D}{C},\latency{O}{D},\latency{D}{O}\}\,,\]
but only
\ile{\{\latency{O}{C},\latency{C}{O}\}}
are quantified here, with the photon trajectories illustrated in \figref{queryToSpacecraft} and \figref{queryFromSpacecraft}.
There are also
two trivial cases \[\latency{O}{D}[\tau] = \latency{D}{O}[\tau] = 2 \distanceTarget/c\]
for any traveler's time $\tau$,
since $\trav{O}$ and $\trav{D}$ are both at rest in $S$.

\incfig
	{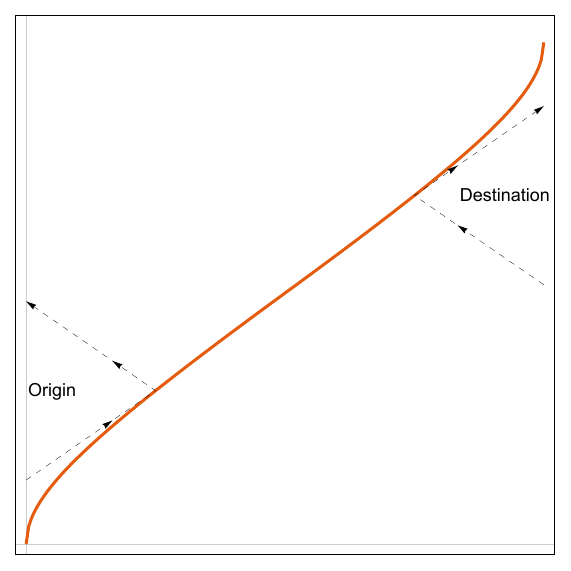}
	{
	trim=0 0 0 0,
   	clip,
    	width=.6\linewidth
	}
	{
An illustration of the photon trajectories for two-way query-response communication
 for a canonical 1-$g$ mission with linear axes.
	Shown are queries originating at the \origin{} and \destination{}, and
	in each case a response from the spacecraft.
	The query-response latencies $\latency{O}{C}$ and $\latency{D}{C}$ 
	are generally very large (multiple years)
	unless the query is generated shortly after launch
	or shortly before landing respectively.
	The query and the response contribute equally to this latency.
}

\incfig
	{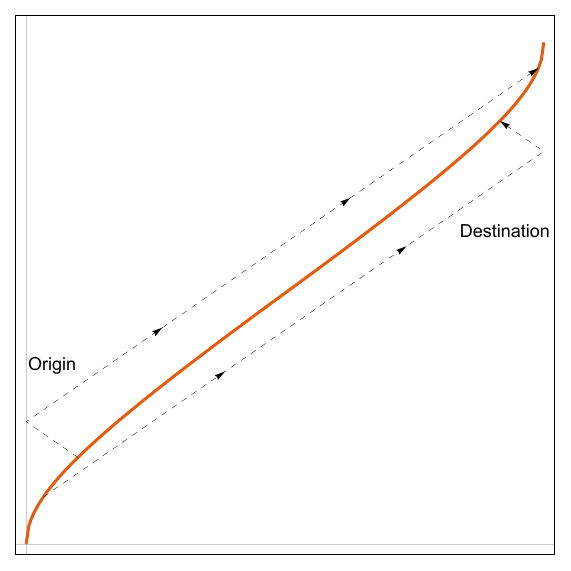}
	{
	trim=0 0 0 0,
   	clip,
    	width=.6\linewidth
	}
	{
A repeat of \figref{queryToSpacecraft} for
	queries from the spacecraft to 
	the \origin{} and to the \destination{}, also with linear axes.
	Again the  latencies $\latency{C}{O}$ and $\latency{C}{D}$ are generally very large (multiple years)
	unless the query is generated shortly after launch
	or shortly before landing respectively.
	The photon propagation in the direction of the spacecraft trajectory
	contributes disproportionally to latency in both cases.
}

$\latency{O}{C}$ is plotted in \figref{responseImageUp} and $\latency{C}{O}$  in \figref{responseImageDown}.
There is a rapid increase in the latency until the point at which the spacecraft begins to slow during the
deceleration phase of the mission.
In the case of $\latency{C}{O}$, that rapid increase actually occurs prior to the event horizon
since the query, following a downlink propagation delay, has to reach $\trav{O}$ prior to the event horizon.

Save for a period following launch on the order of days or months, the conclusion is that
two-way communication becomes at best cumbersome
due to large response latencies
(on the order of years, decades or centuries, except for missions to relatively nearby targets).
Autonomous control of the spacecraft is a must for most of the mission.
The implications are profound for a human crew and for humanity back at the \origin{},
because for most of the mission both must rely 
primarily on one-way 
streaming of crew status and current events (see \secref{imageDef}) rather than a conversation.

\incfig
	{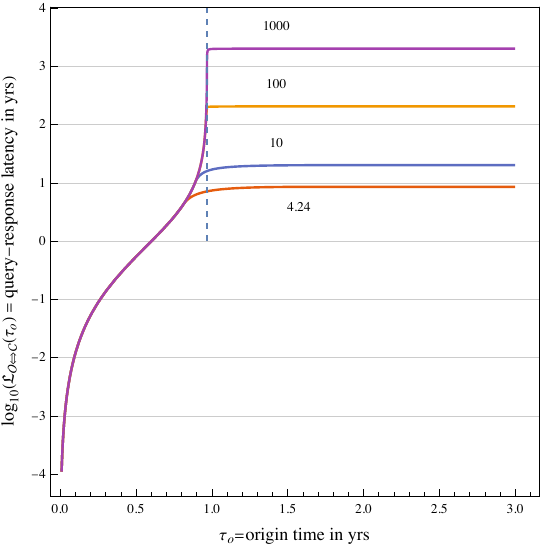}
	{
	trim=0 0 0 0,
    	clip,
    	width=.65\linewidth
	}
	{
A plot of the query-response latency $\latency{O}{C}$ in years
	for queries sent from the \origin{} to the spacecraft
 for a canonical 1-$g$ mission.
	The photon trajectories were illustrated in \figref{queryToSpacecraft}).
The horizontal axis is the time after launch (\ile{\travtime{o} =0}) that a query to the spacecraft is originated
	at the \origin{}, and the vertical axis is the resulting query-response latency.
Only a very small fraction of the total MECT is plotted, especially at the longer distances.
	The vertical axis is logarithmic, reflecting the very large  $\latency{O}{C}$ following 
	the event horizon $\tH$ (shown as a vertical dashed line).
	Queries sent following $\tH$
	result in much greater latency since the query does not reach the spacecraft until the second half of its
	cruise phase. 
	The different curves show $\latency{O}{C}$ for different distances
	\ile{\distanceTarget \in \{4.24,10,100,1000 \} \text{ ly}} to the \destination{},
	with longer distances resulting in greater  $\latency{O}{C}$ beyond the event horizon
	due to the longer duration of acceleration (as opposed to deceleration).
}

\incfig
	{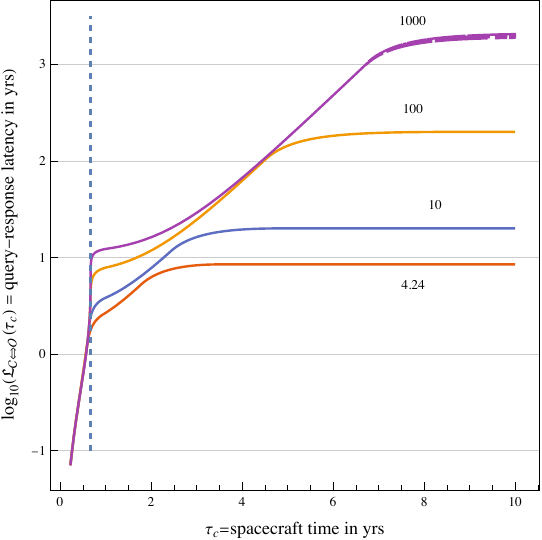}
	{
	trim=0 0 0 0,
    	clip,
    	width=.65\linewidth
	}
	{
A plot of the query-response latency $\latency{C}{O}$ in years
	for queries sent from the spacecraft to the \origin{}
 for a canonical 1-$g$ mission.
	The photon trajectories were illustrated in \figref{queryFromSpacecraft}.
	All the conventions of \figref{responseImageUp}  are repeated, except the
	direction of photon transfer is reversed.
	Thus the horizontal axis is the traveler's time $\travtime{c}$ at which the spacecraft originates the
	query, and the vertical axis is the logarithm of the resulting query-response latency as measured
	by that same traveler's time.
	The vertical dashed line is at
	\ile{\travtime{c}=\tH{\cdot}\log 2}, which is
	the $\travtime{c}$ for which a photon emitted from the spacecraft
	reaches the \origin{} at the event horizon \ile{\travtime{o} = \tH} (see \eqnref{imageConstAccelCO}).
	This is the point at which responses begin to experience much larger delays 
	(see \figref{clockImage1gVsDistanceUp}).
}

\subsection{Time-warping in streaming communication}
\label{sec:streaming}

The query-response latency captures one common type of
communication protocol where one party has a question or
comment and the other party responds.
Our conclusion is that the latency is impractically large
except for communication with the \origin{} for a short period after launch
or with the \destination{} for a short period near landing.
\emph{Streaming} is a second mode of communication, in which
either spacecraft or \origin{} or \destination{} transmits
information continuously with no expectation of a response.
\begin{example}
The \origin{} can transmit a continuous stream of sports
programming to entertain human astronauts (see \secref{streaming}), or the
spacecraft can transmit a webcam image of activity onboard
the spacecraft back to the \origin{}.
\end{example}
Of course achieving sufficiently high data rates to support
reasonable program fidelity is assumed.
(Low data rates may limit the fidelity of the
streamed material.
To improve that fidelity there is the
option to stretch the transmission out in 
time and then transmit it at a lower
data rate.)
Subject to this limitation, streaming is more practical
than query-response communication during the bulk of the cruise
phase in that the inevitable propagation delay
should not disturb the quality of the viewing
experience at the time of reception.

In streaming it is natural to consider a continuous-time transmit and receive signal waveform
$s[\tau]$
representing the signal intensity (average rate of
photon emissions or detections) vs time.
In one common form of modulation,
the underlying data is represented by a
variation of that intensity.
(There are other less energy-efficient modulation techniques that modulate phase and/or wavelength, typically
in the context of heterodyne as opposed to photon-counting detection.)
The effect of accelerated motion is a nonlinear `warp' in the timebase $\tau$ of the waveform $s[\tau]$ between the transmitter and receiver.

When streaming to or from the spacecraft,
it is evident from \figref{trajEnsemble}
(and keeping in mind that coordinate
time is greatly expanded relative to
the spacecraft traveler's time)
that there is a significant expansion or
contraction of a streaming program's duration.
In the case of streaming to or from the \origin{}
(the only case considered here),
a streaming program's duration
is expanded from transmitter to receiver in both directions:
\begin{itemize}
\item
From \origin{} to spacecraft this is due to
the time horizon limitation (which is less than the METT) on total transmit
time.
\item
From spacecraft to origin{} this is due to
the expansion in coordinate time relative to
spacecraft traveler's time
(since \ile{\text{MECT} \gg \text{METT}}).
\end{itemize}

In both cases,
if the program is streamed in ``real time'' at the transmitter,
it is not only delayed (due to propagation delay) but also has an expanded duration at the receiver.
If desired it can be stored at the receiver and subsequently played
back at a higher speed (to return to real time).
This time expansion is due to very different physical effects in the two directions.
For $\direction{O}{C}$ this is due to the rapidly increasing propagation
delay experienced by photons as the event horizon is approached (this is a classical effect).
For $\direction{C}{O}$ this is a manifestation of the relatively slow evolution of
the traveler's time $\travtime{c}$ relative to the coordinate clock $\coordtime{o}$,
which is a manifestation of relativistic time dilation
(this is directly related to the observation that
\ile{\text{METT} \ll \text{MECT}}).

For the streaming of programs of finite length (hours rather than years),
this motion-induced time warp has minimal variation
within the program duration.
The timebase expansion can then be approximated by the derivative of the receiver local time with respect to the
transmitter local time referenced to a specific traveler's time $\travtime{c}$ at the spacecraft,  
and assuming that the result is constant over time intervals of interest.
This derivative is evaluated in \secref{timeWarp} for the
simpler indefinite self-acceleration case, and found to be
\begin{equation}
\label{eq:warpFactor}
\frac{\text d \travtime{c}}{\text d \travtime{o}} \bigg|_{\direction{O}{C}}
= \frac{\text d \travtime{o}}{\text d \travtime{c}} \bigg|_{\direction{C}{O}}
= 
\frac{\psidirection{C}{O}[\travtime{c}]}{\omegadirection{O}{C}[\travtime{c}]}
= \text e^{\travtime{c}/\tH}
\,.
\end{equation}
The timebase is expanded (rather than contracted), and is identical in the two directions,
when referenced to the same instant from the spacecraft perspective,
in spite of the distinctive physical processes at work. 
It increases exponentially with the spacecraft traveler's time $\travtime{c}$
due to the continuous increase in speed under constant acceleration.
\begin{example}
The \origin{} $\trav{O}$ streams
a football match to the spacecraft $\trav{C}$,
and the spacecraft streams back
a webcam video of the astronauts
watching the match so that the controllers
at the \origin{} can gain vicarious enjoyment
from watching the astronauts' reaction
to the goals.

We are interested in the timing of this
interaction.
The appropriate timebases are the
local clocks at $\trav{O}$ and $\trav{C}$
that govern the streaming timebases in the
respective directions.
Assume that these clocks are initialized
so that
\ile{\text{MECT} = \coordtime{o}} at $\trav{O}$
and \ile{\text{METT} = \travtime{c}} at $\trav{C}$.

Consider the case where the match video
arrives at $\trav{C}$ following two 
years of 1-$g$ self-acceleration, or
$\travtime{c}{=}2 \text{ yr}$,
and the webcam video is simultaneously
streamed back.
What are the times of transmission and reception of these streams at the \origin{}?
These times are revealed by the appropriate clock images, which are derived later in \secref{imageQuery},
with the result
\begin{subequations}
\label{eq:exTimes}
\begin{align}
\label{eq:exTimes1}
\direction{O}{C}{:}\ \ 
&\coordtime{o} = \omegadirection{O}{C} \big[\travtime{c}{=}2 \text{ yr}\big] = 0.846 \text{ yr}
= 309 \text{ days}
\\[5pt]
\label{eq:exTimes2}
\direction{C}{O}{:}\ \ 
&\coordtime{o} = \psidirection{C}{O} \big[\travtime{c}{=}2 \text{ yr}\big] = 6.66 \text{ yr}
\,.
\end{align}
\end{subequations}
Although both directions of streaming are concurrent at the spacecraft,
at the \origin{} the $\direction{C}{O}$ webcast is delayed by
\ile{5.8 \text{ yr}}
relative to the $\direction{O}{C}$ match.
The controllers at the origin have to wait multiple years
to observe the astronauts' reaction.

What is the timebase expansion of the two
streaming programs?
Applying \eqnref{warpFactor}, the expansion in $\direction{O}{C}$ happens to be the same as the expansion
$\direction{C}{O}$.
(Again this assumes the spacecraft's reception of the match
and the spacecraft's transmission of the webcast are simultaneous.)
In both cases this expansion is a factor of 7.87.
(The ratio of
\eqnref{exTimes2}
to \eqnref{exTimes1}
also evaluates to 7.87, which is
also anticipated in \eqnref{warpFactor}.)
Thus a one-hour football match would be expanded to 7.87
hours at the receiver.
If the match video is played back as it is received, it would appear to be running in slow motion.
To make the presentation more natural for 
the astronauts, the streamed match could be stored
and subsequently played back at its original speed (which is 7.87 times faster
than it was received).
Identical logic applies independently to the spacecraft's webcast as it is received 
by the controllers.
\end{example}
Similarly, in streaming to and from the \destination{}, it can be shown that the timebase
contracts (rather than expands), is the reciprocal (decreases exponentially), and is identical in the two directions
$\direction{C}{D}$ and $\direction{D}{C}$.
Although these symmetries generalize to an arbitrary spacecraft trajectory,
they are attributable to $\trav{O}$ and $\trav{D}$ both being at rest in $S$ and do not
apply when they are in motion.

\subsection{Communication between two spacecraft}
\label{sec:twoRockets}

We can imagine two spacecraft with the same
acceleration profile, but launched at different times,
for example in an alien-world colonization scenario.
In that case it is interesting to consider the possibilities for communication between 
this pair of spacecraft.

Suppose that there are two spacecraft $\trav{C1}$ and $\trav{C2}$
launched in that order, so the ordering of spacecraft versus distance is
\[ \trav{O} \rightarrow \trav{C2} \rightarrow \trav{C1} \rightarrow \trav{D} \,.\]
Thus during their respective cruise phases, their trajectories are illustrated in \figref{trajTwoRockets}.
The traveler's times aboard the two spacecraft initialized to \ile{\travtime{c1}=\travtime{c2}=0} at their respective launch times,
so that they measure the METT for each spacecraft.

Our immediate concern is exploration of the photon exchange between
 \ile{\{\trav{C2}, \trav{C1}\}} in both directions.
 As illustrated in \figref{trajTwoRockets}, photon exchange $\direction{C2}{C1}$
  is feasible as long as the launch time of
$\trav{C2}$ falls within \ile{0<\coordtime{o}<\tH} (as it does in this case).
However, the photon emission time is limited by the event horizon of $\trav{C1}$, after which there
is a communication blackout.

 \incfig
	{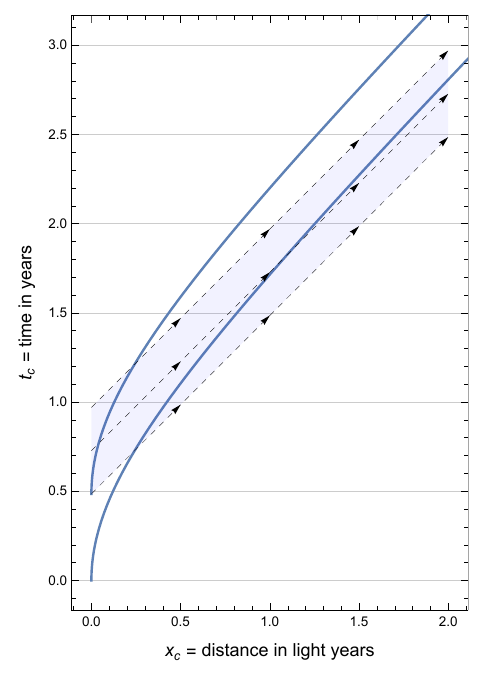}
	{
	trim=0 0 0 0,
    	clip,
    	width=0.5\linewidth
	}
	{
A plot of the trajectories of two spacecraft launched from the \origin{} $\trav{O}$ at different times,
	each with indefinite 1-$g$ self-acceleration.
	One spacecraft (which is traveler $\trav{C1}$) is launched at \ile{\coordtime{o}=0}
	and the other (which is traveler $\trav{C2}$) is launched at  \ile{\coordtime{o}=0.5 \cdot \tH}.
	Also shown as dashed lines are the trajectories of a photon emitted from the \origin{}
	at three coordinate times \ile{\coordtime{o} \in \{0.5 ,0.75, 1 \} \cdot \tH} in the $+x$ direction.
The shaded area represents the ensemble of photon emission times from $\trav{C2}$ which
	intersect the trajectory of $\trav{C1}$ (and hence can be detected by $\trav{C1})$.
	This represents a finite interval beginning at the instant of launch of $\trav{C2}$.
}

The clock images for communication in both directions are illustrated in 
\figref{twoRocketImageUp} and \figref{twoRocketImageDown}
for a $\trav{C2}$ launch coordinate time  \ile{\coordtime{o}=\epsilon \tH} where \ile{0<\epsilon<1}
(see \secref{imageTwo}).
The query-response latencies for two-way communication between $\trav{C2}$ and $\trav{C1}$ are
illustrated in \figref{twoRocketLatencyUp} and \figref{twoRocketLatencyDown}.

\incfig
	{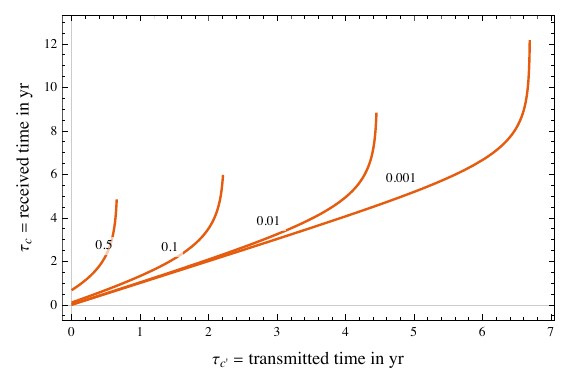}
	{
	trim=0 0 0 0,
    	clip,
    	width = 0.7\linewidth
	}
	{
The clock image $\psidirection{C2}{C1}$ for photon exchange from $\trav{C2}$ to $\trav{C1}$ is plotted.
	Two spacecraft $\trav{C1}$ and $\trav{C2}$ are launched in that order,
  so that
	photon propagation is in the same direction as the spacecraft trajectories.
	Both spacecraft are assumed to have indefinite
	1-$g$ acceleration,
and the traveler's times \ile{\{\travtime{c1},\travtime{c2}\}}
      each measure METT for their respective spacecraft
     (\ile{\travtime{c1}=\travtime{c2}=0} at launch).
The different curves represent \ile{\epsilon \in \{0.1,0.5,0.9 \}}, where 
	the later-launch is delayed by $\epsilon\,\tH$ relative to the earlier launch.
	The time $\travtime{c1}$ for photon detection by $\trav{C1}$ occurs increasingly late in its trajectory
	as $\epsilon$ increases.
	There is a limit \ile{-\,\tH \cdot \log \epsilon} on the emission time $\travtime{c2}$ that falls within
	the time horizon of $\trav{C1}$.
	The image $\psidirection{C2}{C1}$ is unbounded toward the end of that interval.
}
	
\incfig
	{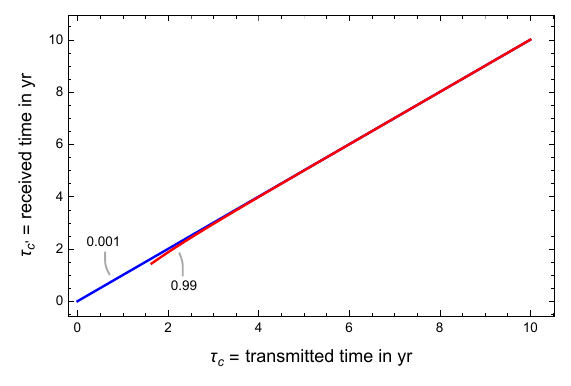}
	{
	trim=0 0 0 0,
    	clip,
    	width = 0.65\linewidth
	}
	{
\figref{twoRocketImageUp} is repeated for photon exchange from an earlier-launched $\trav{C1}$ spacecraft
	to a later-launched spacecraft $\trav{C2}$, with
	photon propagation in the opposite direction to the spacecraft trajectories.
	The clock image is $\psidirection{C1}{C2}$ and
	for this case a communication blackout never occurs.
	The earliest photon emission time $\travtime{c1}$ by $\trav{C1}$ is
	\ile{\tH \cdot \log (1+\epsilon)}, since earlier emissions would arrive prior to the launch of $\trav{C2}$
 (recall that $\trav{C2}$ is launched later
 than $\trav{C1}$).
	Later in $\trav{C1}$'s trajectory, $\epsilon$ has minimal impact on  $\psidirection{C1}{C2}$.
}

\incfig
	{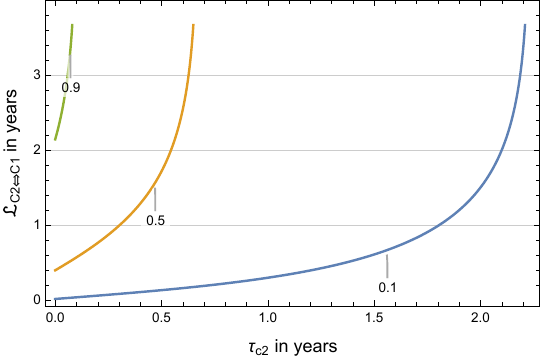}
	{
	trim=0 0 0 0,
    	clip,
    	width = 0.65\linewidth
	}
	{
The query-response latency $\latency{C2}{C1}$ for a query initiated by
	the later-launched rocket $\trav{C2}$ to the earlier-launched rocket $\trav{C1}$ is plotted
 for the same assumptions as in \figref{twoRocketImageUp}.
	Queries can be initiated by $\trav{C2}$ immediately following launch,
	but receiving a response is only possible for an interval 
	\ile{0 < \travtime{c2} <-\,\tH \cdot \log \epsilon} following launch.
	Toward the end of that interval $\latency{C2}{C1}$ becomes unbounded.
}

\incfig
	{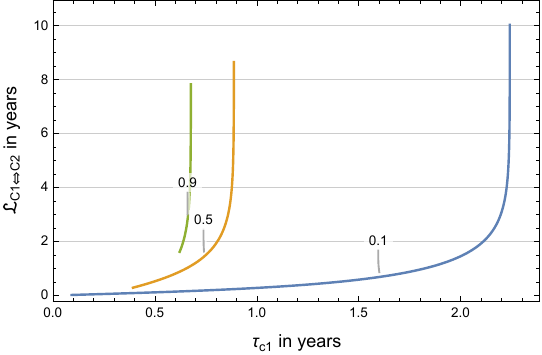}
	{
	trim=0 0 0 0,
    	clip,
    	width = 0.65\linewidth
	}
	{
	The query-response latency $\latency{C1}{C2}$ 
	for a query initiated by
	the earlier-launched spacecraft $\trav{C1}$.
 This is a repeat of \figref{twoRocketLatencyUp} for
 a query in the opposite direction.
Both the earliest and latest such query is
 constrained.
	The earliest query arrives at $\trav{C2}$ at the instant of its launch.
	There is also a latest such query time  
 \ile{\tH \cdot \log \left( \epsilon + 1/\epsilon \right)} 
 that allows for a response from $\trav{C2}$
 to reach $\trav{C1}$.
This is because that response must be initiated early enough
to fall within the
	time horizon of $\trav{C1}$.
	As the time horizon is
 reached, $\latency{C1}{C2}$
    grows without bound.
}

\subsection{Role of acceleration magnitude and duration}
\label{sec:durationAccel}

The canonical mission of \figref{launchLandingMission} makes the assumption that
acceleration magnitude is time-invariant during the entire travel time from launch to landing.
Having developed a complete picture of one-way image and two-way query-response latency,
two modifications that can be made in the interest of easing the restrictions 
on communication can be considered.
One is reducing the acceleration and deceleration time durations,
thereby introducing an intermediate `ballistic' interval during which the spacecraft speed is constant.
The second is reducing the acceleration magnitude 
$\alpha$
(thereby increasing $\tH$).
Neither will make a meaningful change in MECT, 
and hence neither change would
have significant impact on the \origin{}'s awareness of landing, nor on
response latencies.
However, these modifications would have a significant effect on one-way uplink communication
to the spacecraft by lengthening the event horizon $\tH$
(allowing longer communication with the \origin{} during the acceleration phase)
and/or by earlier termination of the acceleration phase
(allowing uplink signals transmitted past the event horizon to be received earlier).
Employing a combination of larger $\tH$ and a shorter acceleration phase
could allow an extension of the event horizon past the acceleration phase altogether.

A significant impact of either of these modifications would be in increasing the
METT,
with the resulting adverse implications for resources aboard the spacecraft and the
required lifetime of equipment and crew.
The greater mass associated with greater resources would be compensated
by lower mass devoted to spacecraft propulsion.
The human crew would devote more of their lifetime to the cruise phase
(and commensurately less to the post-landing) experience.

\section{Analysis}
\label{sec:analysisRelativity}

Relativistic effects can generally be neglected in the case of
spacecraft missions within our Solar System, at least when chemical
spacecraft propulsion is
employed and the speeds are relatively low compared
to light-speed $c$.
This implies that the Newtonian theory of gravity and
a single universal measure of coordinate time
in the predication of spacecraft
trajectories and communication latencies is suitably accurate.
This approach has to be abandoned
for interstellar missions, where the full General Theory of Relativity (GR) 
\cite{RefNumber675} could be adopted.
In this paper the `empty-space' approximation is utilized, in which the
spacecraft trajectory is determined entirely by its own propulsion, neglecting
the effects of gravity from astronomical bodies including the \origin{} and \destination{}.
This is the standard approach for development of the relativistic spacecraft equation 
\cite{bade1953relativistic,forward1995transparent},
and relies on the simpler (and more analytically tractable)
Special Theory of Relativity (SR).

There are numerous resources for studying SR, including some excellent
textbooks \cite{RefNumber705,RefNumber704,RefNumber703}.
We follow an alternative approach and notation to the development of SR
originating in \cite{RefNumber806}, which (it was argued) is more appropriate for
engineering applications such as interstellar travel (as opposed to scientific applications).
In particular the perspective of the traveler (accelerated observer) was adopted in place
of the usual denizen (inertial) observer universally adopted in physics, starting with
Einstein's initial formulation.
This alternative development greatly simplifies the derivation of trajectories involving acceleration,
and is consistent with the universal practice of relying on clocks carried by spacecraft.
This approach was inspired in part by earlier pedagogical innovations by
Fraundorf \cite{RefNumber693,RefNumber694,RefNumber692}
and a new derivation of the relativistic spacecraft equation by Walter \cite{RefNumber674,westmoreland2010note}.
The notion of a clock image, which forms the basis of the communication latencies
developed here, was introduced in \cite{RefNumber806}.

Most relativistic spacecraft derivations assume that the propulsion is
based on a fixed-rate consumption of a propellant carried by the spacecraft,
in which case the force exerted on the spacecraft will be constant.
As the mass of the spacecraft decreases, the
self-acceleration thereby increases.
In contrast,
the canonical mission considered here assumes that the self-acceleration of the spacecraft is fixed.
This would require that the rate of fuel consumption start out high to compensate
for the higher initial fuel mass,
and decrease during the mission.
Alternative versions of the spacecraft equation are derived and compared in \cite{RefNumber806}.
Here we do not take propulsion challenges into account, but rather make the 
`empty-space constant self-acceleration' simplifying assumption
in order to focus our efforts on understanding the communication timing issues
in a simple context.
Our methods readily extend to more complicated mission scenarios.

\subsection{Map and two clock metaphor}

A map and two-clock metaphor \cite{RefNumber806} forms the basis for the following.
The kinematic variables used in this metaphor are summarized in \tblref{metaphor}.

\doctable
	{metaphor}
    	{Kinematic variables for a 1-D map and two-clock metaphor}
   	{lll}
   	{
\\ \hline \\ [-1ex]
Variable & Traveler's version & Coordinate version
\\ \hline \\ [-1ex]
Map-position & $r$ & $x$
\\[5pt]
Time & $\tau$ & $t$
\\[2pt]
Time-speed 
&& $\displaystyle\gamma = t^\prime =  \frac{\text d t}{\text d \tau}$
\\[10pt]
Map-speed & $\displaystyle w = r^\prime = \frac{\text d r}{\text d \tau} $ & $ \displaystyle u = x^\prime = \frac{\text d x}{\text d t}$
\\[10pt]
Self-acceleration & $\alpha$
\\[5pt]
Map-acceleration 
& $\displaystyle w^\prime = \frac{\text d w}{\text d \tau} $ 
&  $\displaystyle u^\prime = \frac{\text d u}{\text d t} $
\\[10pt] \hline
      }

A coordinate system to describe the trajectory of a moving body like a spacecraft $\trav{C}$ is
defined relative to an inertial frame $S$ which defines a set of spatial coordinates.
Relative to $S$, any observer  $\trav{R}$ with fixed spatial coordinates (said to be \emph{at rest})
experiences no acceleration.
The coordinates defined by $S$ are called a \emph{map}, in analogy to terrestrial travel,
and the position of an observer is called the \emph{map-position}.
For the general case a map-position is three-dimensional, but for simplicity here
we specialize to the one-dimensional (rectilinear) case.
Thus there is a single scalar spatial coordinate, which is labeled $r$ or $x$.
Any body such as $\trav{C}$ not permanently at rest is called a traveler.

There are two types of clocks of interest.
The traveler's clock (carried along with $\trav{C}$),
is the clock of interest in the context of interaction (such as communication) among two or more
travelers, and measures traveler's time $\tau$ (as was described in \secref{localClocks}).
Any clock at rest in $S$ is called
a coordinate clock,
because it measures \emph{coordinate time} denoted by $t$.
Traveler and coordinate clocks advance at different rates, excepting the degenerate
case of a traveler temporarily or permanently at rest relative to $S$. 
Clock synchronization,
the issue of initialization of all clocks, was discussed in \secref{clockSynch}.

Kinematic variables quantify the motion of travelers, and for our purposes
constitute map-position as well as the map-speed and map-acceleration measured at the same map-position.
SR offers a complete model of kinematics, for accelerated bodies as well as inertial, in the absence of gravitational influence.
Since map-speed is the derivative of map-position, and map-acceleration is the
derivative of map-speed, both relative to an appropriate definition of time,
there are two distinct sets of kinematic variables.
The traveler's set, measured relative to $\tau$, is denoted by \ile{\{\tau,r,w,w^\prime\}},
where $w^\prime$ denotes the derivative of $w$ with respect to $\tau$.
Similarly the coordinate set \ile{\{t,x,u,u^\prime\}} is measured relative to $t$.

In addition to  \ile{\{w^\prime,u^\prime\}},
there is a third measure of acceleration $\alpha$ measured by
an accelerometer carried aboard the spacecraft,
and called the \emph{self-acceleration}.
Any accelerometer must implicitly incorporate a measurement of time,
and the appropriate clock for this purpose is the traveler's clock
(which is carried along with that accelerometer as part of the traveler's payload).
Thus $\alpha$ is unambiguously measured relative to traveler's time $\tau$
(there is no coordinate-time version).
In general the three forms of acceleration \ile{\{w^\prime,u^\prime,\alpha\}}
possess different numerical values at any
instant of time.
The coordinate map-acceleration $u^\prime$ is of no concern in the following.

Finally, the \emph{time-speed} $\gamma$ 
(called the \emph{Lorentz factor})
measures the rate of advance of a coordinate clock $t$
relative to a traveler's clock $\tau$ 
situated at the same position.

\subsection{Time dependence of kinematic variables}
\label{sec:SRreview}

Since there are two distinct time measurements of interest,
to keep things straight we exercise care
in expressing the time dependence of kinematic variables.
There are two kinematic variables  that are unaffected by the choice of $S$,
and depend only on $\tau$ and not $t$:
These are $\tau$ itself and self-acceleration $\alpha[\tau]$.
The other kinematic variables not only have a coordinate-time version, but more fundamentally
these variables are affected by the choice of $S$.
Since it is frame-invariant, $\tau$ is preferred as the fundamental timebase for accelerated motion,
and the traveler's time kinematic variables are written as
 \ile{\{\tau,\gamma[\tau],r[\tau],w[\tau],w^\prime[\tau]\}}.
 This kinematic description is also practical because a traveler is assumed to carry a clock measuring $\tau$,
 and the traveler is normally navigationally aware of its own position $r[\tau]$.
 
 This is so simple and clean, it is reasonable to ask ``in the communications context, what purpose is served by a coordinate-time
 description of kinematics?''
 A fundamental postulate of SR is that the speed of light $c$ is invariant to the choice of inertial frame $S$.
 The interaction among different travelers is based on
 photon exchange, and the propagation delays for photons are defined in terms
 inertial frame $S$ and coordinate time $t$.
 Thus for us to quantify photon-transfer interactions, for each traveler emitting or detecting
 a photon we must know its traveler's time $\tau$ and map-position $r[\tau]$ at that instant,
 but also the associated coordinate time at that map-position and instant of time,
 which we write as $t[\tau]$.
 It would be unusual for there to
 be an actual physical clock at rest at that map-position, so typically this coordinate clock is \emph{notional}
 rather than physical.
 The entire set of kinematic variables can thus be written as 
 \ile{ \big\{ t[\tau], x\big[t[\tau]\big], u\big[t[\tau]\big], u^\prime \big[t[\tau]\big] \big\} }.
 For purposes of photon transfer timing calculations (the focus of this paper),
 the only coordinate variable needed is $t[\tau]$, since map-position is conveniently specified by
 $r[\tau]$.
 
 The calculations establishing all the useful kinematic variables
 proceeds in the following sequence:
 \begin{subequations}
 \label{eq:orderCalc}
 \begin{align}
 \label{eq:orderCalc1}
 &\alpha[\tau] \rightarrow w^\prime[\tau] \rightarrow w[\tau] \rightarrow r[\tau]
 \\
 \label{eq:orderCalc2}
&  w[\tau] \Rightarrow \gamma[\tau] \rightarrow t[\tau]
\,.
\end{align}
\end{subequations}
Each `$\rightarrow$' represents the solution to a first-order differential equation (DE) given in \tblref{metaphor}
(with the exception of \ile{\alpha[\tau] \rightarrow w^\prime [\tau]}, which is noted below),
and the single `$\Rightarrow$'  is a functional relation also noted below. 
All kinematic variables follow from knowledge of $\alpha[\tau]$, which is the input.
The final outcome \ile{\{ r[\tau], t[\tau] \}} is everything needed for the photon transfer propagation delay calculations
that enter into clock images and the resulting query-response latencies.

\subsection{Kinematic variable solutions}

The role of time-speed $\gamma[\tau]$ becomes evident when
the two versions of map-speed are compared.
Applying the chain rule of differentiation,
\begin{equation}
\label{eq:speedTravelers}
w[\tau] = r^\prime [\tau]
= x^\prime \big[t[\tau]\big] \cdot t^\prime [\tau] 
= u\big[t[\tau]\big] \cdot \gamma [\tau]
\,.
\end{equation}
Thus $\gamma[\tau]$ directly relates the two map-speeds.
The \ile{w[\tau] \Rightarrow \gamma[\tau]} relation in \eqnref{orderCalc2} is \cite{RefNumber806}
\begin{equation}
\label{eq:Lorentz}
\gamma[\tau] = \sqrt{1 + w[\tau]^2 / c^2}
\,.
\end{equation}
If $S$ is changed for any reason, both $\gamma[\tau]$ and $w[\tau]$ are affected as a result, but
their frame-invariant relationship in \eqnref{Lorentz} is preserved.

Relation \eqnref{Lorentz} has profound implications for interstellar travelers
(analogous to \ile{E = m c^2} for energy and propulsion)
because it directly relates motion and time.
In particular, since for a moving body \ile{1 < \gamma[\tau]}, it follows that
$\tau$ always advances more slowly than $t[\tau]$,
and the larger the map-speed, the larger this \emph{time-dilation} effect.
A natural question is ``what is the asymmetry that causes this disparity to always occur in the same direction''.
One simple answer is that a measurement of the traveler kinematic variables expressed in terms of
traveler's time can be accomplished using
the single traveler's clock carried by the traveler, 
but measuring such variables expressed in terms of coordinate time requires two or more
coordinate clocks.

\eqnref{Lorentz} and \eqnref{speedTravelers} together imply that (\ile{| u\big[ t[\tau] \big] | < c}),
leading to the common perception that objects cannot travel faster than light.
This is at best an incomplete statement when applied to interstellar travel, 
because spacecraft self-aware of their own motion employing their own clock
as the basis of observations see no such limitation on speed
(since \ile{|w| \gg c} is possible).
The spacecraft's perception of super-luminance (travel faster than light-speed) is
attributable to its traveler's clock advancing more slowly.
There is no inconsistency since a traveler's clock is incapable of measuring
photon speed.
Since \ile{| u | < c} a traveler's clock as observed in $S$
necessarily travels slower than a photon, and thus could measure either the time of a
photon's emission or the time of that same photon's detection, but never both.

The \ile{\alpha[\tau] \rightarrow w^\prime[\tau]} DE in \eqnref{orderCalc1} is given by
\cite{RefNumber806}
\begin{equation}
\label{eq:selfToMapAccel}
w^\prime [\tau] = \gamma [\tau] \cdot \alpha [\tau] \,.
\end{equation}
After substituting for $\gamma[\tau]$ from \eqnref{Lorentz}, 
this becomes a first-order
DE in $w[\tau]$.
This is another frame-invariant relation; that is, both $w[\tau]$ and $\gamma[\tau]$ are
affected by the choice of $S$, but relationship \eqnref{selfToMapAccel} is not.
The self-acceleration $\alpha[\tau]$ is of practical interest for two important reasons:
\begin{itemize}
\item
Acceleration $\alpha$ relates directly to the force imparted by the propulsion system \cite{RefNumber674}, 
if propulsion is the cause of the acceleration.
(Another possibility is gravity, but that effect is neglected in SR.)
\item
Acceleration $\alpha$ is that experienced by on-board equipment and material, including biological material,
both of which are normally sensitive to and affected by the acceleration magnitude.
In numerical examples a time-invariant 1-$g$ 
self-acceleration magnitude is assumed, because this is consistent with that normally
experienced by humans and other biological material originating from Earth.
\end{itemize}

Since again \ile{1<\gamma[\tau]}, what follows from \eqnref{selfToMapAccel} is the profound observation that
\ile{| w^\prime [\tau] | > | \alpha [\tau] |}, which is called an \emph{acceleration boost}.
That map-acceleration is always greater than self-acceleration
is great news for interstellar travelers.
Although biological organisms (like humans) have a limited tolerance for
high self-acceleration, that intolerance need not limit the map-acceleration.
Even better news for interstellar travelers is a positive
feedback effect:
As $w[\tau]$ grows so does $\gamma[\tau]$, and hence acceleration boost magnitude is amplified
without limit for as long as self-acceleration (in the direction of travel) persists.
This makes it advantageous to maintain self-acceleration as long as possible
(as in the canonical mission assumption, see \secref{durationAccel}), because longer self-acceleration reduces the METT
(while its effect on MECT is minimal)
and thus reduces the resources (food, water, oxygen, etc) required on the spacecraft, as well as biological aging.

\subsection{Trajectory with indefinite self-acceleration}
\label{sec:trajectoryIndefinite}

The coordinate-time spacecraft trajectory \ile{\{ r_c [ \travtime{c}], \coordtime{c}  [ \travtime{c}] \}}
for constant acceleration and time horizon $\tH$
(see \secref{trajConstAccel}) is
 \begin{subequations}
 \label{eq:motionVarsIndefiniteAccel}
\begin{align}
\label{eq:motionTimeCoordTime}
&\coordtime{c}[\travtime{c}] = \tH \cdot \sinh \big(\travtime{c}/\tH \big)
\\
\label{eq:motionPositionCoordTime}
&r_c[\travtime{c}] = c\, \tH \Big( \cosh \big(\travtime{c}/\tH \big) - 1 \Big)
\,.
\end{align}
\end{subequations}
This is consistent with the textbook \emph{hyperbolic trajectory} for a uniformly
accelerated body (see Ch.14 of \cite{RefNumber703}), although
here it has been derived in an unconventional way.
\begin{example}
Let the units of time be year and the units of distance be light years, and
determine the kinematic variables 
at \ile{\travtime{c}{=}2}.
The coordinate map-speed is \ile{u{=}0.968} (96.8\% of $c$), the traveler's map-speed is \ile{w{=}3.87}
(387\% of $c$),
and the time-speed is \ile{\gamma{=}4.0}.
While the traveler's self-acceleration is \ile{\alpha = g{=}1.03}, the traveler's map acceleration
is \ile{w^\prime{=}4.13} (\ile{\gamma{=}4.0} times the self-acceleration $\alpha$).
\end{example}

\subsection{Event horizon}
\label{sec:tH}

Before verifying the event horizon analytically, the interpretation of the phrase
``trajectory of a photon'' should be clarified.
As \ile{u\big[t[\tau]\big] \to c} we find that \ile{\gamma[\tau] \to \infty}, which implies that \ile{\tau \to 0}.
Thus, from the perspective of a photon (if indeed it had its own perspective!) we have \ile{\tau \equiv 0}.
A photon trajectory cannot be parameterized by $\tau$ (as in the case of non-zero mass particles).
Rather, a photon trajectory \ile{\{x[t],t\}} must omit $\tau$ altogether.
A fundamental postulate of SR is that a photon's position versus coordinate time
obeys the \emph{range equation}
\begin{equation}
\label{eq:range}
t = t_0 + x[t]/c
\end{equation}
for a photon that is emitted from position \ile{x[0]=0} at time \ile{t = t_0}.
This expresses the simple concept that the photon travels at speed $c$ 
from the perspective of observers at rest in inertial frame $S$.
Photon trajectories in previous plots have this special interpretation.

Analytically verifying the time horizon phenomenon follows from comparing two coordinate times:
\begin{description}
\item[Spacecraft.]
When the spacecraft travel time is $\travtime{c}$, its distance reached
is \ile{r_c [ \travtime{c} ]} and
the coordinate time at which it reaches this distance
is \ile{\coordtime{c}[\travtime{c}]}.
\item[Photon.]
Based on \eqnref{range}, the coordinate time that a photon emitted from the \origin{}
at coordinate time 
\ile{\coordtime{c}[\travtime{c}] = \tH} 
arrives at the same distance \ile{r_c [ \travtime{c} ]} is
\ile{\big(\tH{+}r_c[\travtime{c}]/c\big)}.
\end{description}
Substituting from \eqnref{motionVarsIndefiniteAccel},
the difference between these two times
\begin{equation}
\big(\tH + r_c[\travtime{c}]/c\big) -\coordtime{c}[\travtime{c}] = \tH \cdot \text e^{-\travtime{c}/\tH} >0
\end{equation}
is positive and approaches zero exponentially as \ile{\travtime{c}{\to}\infty}.
Thus the photon and accelerated spacecraft trajectories approach one another (but never intersect).
The event horizon no longer applies when
the spacecraft terminates its acceleration and/or starts decelerating,
as illustrated in \figref{trajCanonicalMission}.

\incfig
	{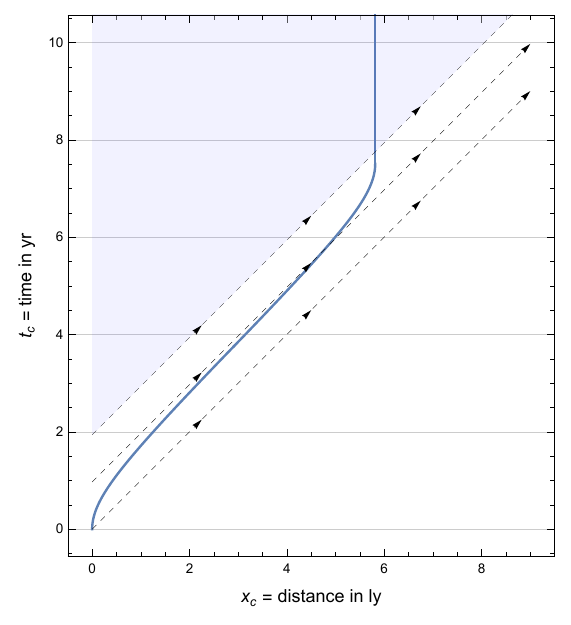}
	{
	trim=0 0 0 0,
    	clip,
    	width=0.6\linewidth
	}
	{
 Illustration of the trajectory of a canonical launch-to-landing 1-$g$ mission as defined in \figref{notation}.
 The \origin{} to \destination{} distance
 $\distanceTarget$ is chosen so that the
 total METT at landing is \ile{\travtime{c} = 4 \text{ yr}}
	(see \secref{trajectoryCanonical}).
	This is a repeat of \figref{trajConstAccel} with the
 addition of a deceleration cruise phase.
	The dashed lines show the trajectories of photons emitted from the \origin{} at \ile{\coordtime{o} \in \{ 0, \tH, 2 \tH \}}.
In this case the spacecraft can detect photon emissions during the interval 
 \ile{\sim \tH < \coordtime{o} < \sim 2 \tH} during the
 deceleration, but emissions for
 \ile{\sim  2 \tH < \coordtime{o}} can only be
 detected following landing at the
 \destination{}.
 A photon that is emitted from
	the \origin{} at \ile{\coordtime{o}=2 \tH} is detected
	shortly following landing,
	so that \ile{\text{MECT} \sim ( D/c + 2 \tH )}.
}

\subsection{Clock images and query-response latencies}
\label{sec:imageQuery}

There are two clock image functions (see \secref{imageDef}) of interest,
$\psidirection{O}{C}$ and 
$\psidirection{C}{O}$.
For trajectory \eqnref{motionVarsIndefiniteAccel} (see \secref{imageAnal}),
\begin{subequations}
\label{eq:imageConstAccel}
\begin{align}
\label{eq:imageConstAccelOC}
&\travtime{c} = \psidirection{O}{C} [\travtime{o}]= -\, \tH \cdot \log \big( 1 - \travtime{o}/\tH \big) \,,\ 0 \le \travtime{o} < \tH
\\
\label{eq:imageConstAccelCO}
&\travtime{o} = \psidirection{C}{O} [\travtime{c}] =  \tH \big( e^{\travtime{c}/\tH} - 1  \big) \,,\ 0 \le \travtime{c}
\,.
\end{align}
\end{subequations}
The domain of $\psidirection{O}{C}$ is restricted to the interval between launch and event horizon
as expected,
whereas the domain of $\psidirection{C}{O}$ is only restricted to post-launch,
although it grows exponentially.
The corresponding inverse clock image functions are
\begin{subequations}
\label{eq:imageFns}
\begin{align}
\label{eq:imageFns1}
&\travtime{o} = \omegadirection{O}{C}[\travtime{c}] = \tH \cdot \left(  1 - \text e^{-\travtime{c}/\tH} \right) \,,\ 0 \le  \travtime{c}
\\
\label{eq:imageFns2}
&\travtime{c} = \omegadirection{C}{O}[\coordtime{o}] =  \tH \cdot \log \left( 1 + \travtime{o}/\tH \right) \,,\ 0 \le  \travtime{o}
\,.
\end{align}
\end{subequations}
The resulting query-response latencies are (see \secref{imageAnal}),
 \begin{subequations}
  \label{eq:latencyFixedAccel}
  \begin{align}
  &\latency{O}{C}[\travtime{o}] =  \frac{\travtime{o}^{\,2}}{\tH - \travtime{o} } \,,\ 0 \le \travtime{o} < \tH
  \\
 &\latency{C}{O}[\travtime{c}] =  -\, \tH \cdot \log \big( 2 - \text e^{\travtime{c} / \tH} \big) - \travtime{c} \,,\ 0 \le \travtime{c} < \tH \cdot \log 2 \,.
  \end{align}
  \end{subequations}
 The range of $\latency{C}{O}$ is more restricted (since \ile{\log 2 \approx 0.69}) because the
 query has to reach the \origin{} early enough that the \origin{}'s response falls within the time horizon of the spacecraft.

\subsection{Other motion effects}

Closely related to the warping phenomenon is red Doppler shift,
which becomes substantial at relativistic speeds
(not considered further here).
This (as well as relativistic aberration) affects the directivity of transmit and/or receive apertures, detectors, etc.
On longer missions multiple receivers are needed to deal with
a wide variation in received wavelength,
or alternatively the transmitter has to employ multiple compensatory wavelengths
to hold the receive wavelength in a narrower range.

Due to both red shift and timebase expansion, the signal power decreases substantially from transmitter to
receiver.
Power and energy are conspicuously not conserved when they are compared with respect to two distinct frames of reference,
the \origin{} and the spacecraft.
Reduced signal power substantially reduces the data rate and data volume that can be achieved,
and this effect increases as long as the constant self-acceleration continues.

\section{Canonical mission trajectory}
\label{sec:trajectoryCanonical}

The kinematic variables in the specific context of the canonical mission scenario
(defined in \secref{missionCanonical})
are illustrated in \figref{notation}.
Although the trajectory of a spacecraft $\trav{C3}$ on a canonical mission
could be calculated directly from the modeling DEs given by SR,
it is easier to recognize the symmetry of canonical self-acceleration.
In particular, the trajectory for the second half of the cruise phase
is related to the trajectory in the first half 
(see \secref{trajectoryIndefinite})
with time running backward,
from maximum to zero map-speed (rather than the other way around).
The second half of the $\trav{C3}$ trajectory is thus the mirror image of the first half of the $\trav{C1}$ trajectory.

\incfig
	{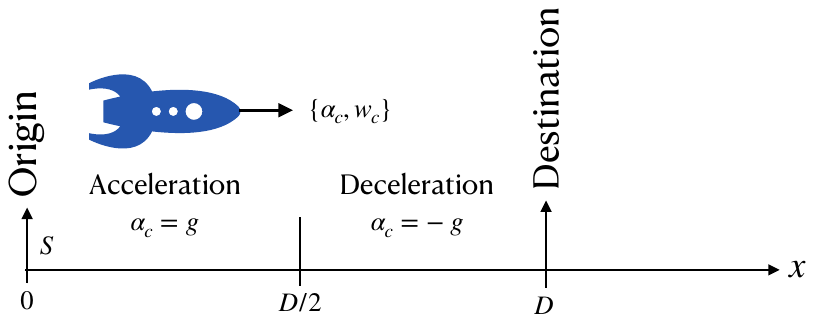}
	{
	trim=0 20 370 350,
    	clip,
    	width=0.6\linewidth
	}
	{
A spacetime coordinate system 
	and the kinematic variables describing the trajectory of a spacecraft relative to
	an inertial frame $S$ with coordinates defined in \figref{trajectory}.
	Added variables are the self-acceleration $\alpha_c$ and traveler's map-speed $w_c$,
	which are both implicitly a function of traveler's time $\travtime{c}$.
	For the canonical 1-$g$ launch-landing mission of \figref{launchLandingMission},
	the acceleration has only two constant values of acceleration,
	\ile{\alpha_c = \pm g}.
}

To determine that mirror image the two times
\ile{\{\travtime{h},\coordtime{h}\}} at the halfway point (\ile{r_c[\coordtime{h}] = \distanceTarget/2}) are needed.
These are determined from \eqnref{motionVarsIndefiniteAccel} to be
\begin{equation}
\label{eq:halfwayTimes}
\travtime{h} = 
\tH \cdot \cosh^{-1} \left( 1 + \frac{\distanceTarget}{2 c \cdot \tH} \right)
\ \ \ \text{and}\ \ \ 
\coordtime{h} = 
\frac{D}{2 c} \sqrt{1 + \frac{4 c \cdot \tH}{D}}
\,.
\end{equation}
It can be verified that
\ile{\coordtime{h} \rightarrow \big( \tH + \distanceTarget/2c \big)}
as 
\ile{\distanceTarget \to \infty},
or in words $\coordtime{h}$ at longer distances approaches the coordinate time at which a photon
emitted from the \origin{} at \ile{\coordtime{o} = \tH} reaches the halfway point.

The second half of the canonical trajectory equals the first half
reversed in time, but in addition displaced
in both coordinate time (by $\coordtime{h}$)
as well as in map-position (by $\distanceTarget/2$).
Taking these three factors into account, the canonical trajectory becomes
\begin{equation}
\{ \coordtime{c3} [\travtime{c3}] , r_{c3} [\travtime{c3}] \} = 
\begin{cases}
\{ \travtime{c3} ,0\}
\,,\ \ 
\travtime{c3} < 0
\\
\{ \coordtime{c1} [\travtime{c3}], r_{c1} [\travtime{c3}] \} 
\,,\ \ 
0 \le \travtime{c3} \le \travtime{h}
\\
\{ 
2 \coordtime{h}- \coordtime{c1} [2\, \travtime{h} - \travtime{c3}], 
D - r_{c1} [ 2 \travtime{h} - \travtime{c3} ]
\}
\,,\ \ 
\travtime{h} < \travtime{c3} \le 2 \travtime{h}
\\
\{ 2 \coordtime{h} + \travtime{c3} - 2 \travtime{h}, \distanceTarget \}
\,,\ \ 
 2 \travtime{h} < \travtime{c3}
\,.
\end{cases}
\end{equation}

\section{Conclusions}

The timing relationship (characterized by the clock image functions, in terms of local traveler's clocks) 
between photon emission and detections as measured at
the \origin{} of launch, and one or two accelerating spacecraft, and a landing \destination{} has been characterized
by drawing on SR.
The methods employed apply generally to accelerated bodies performing photon transfer in their communications.
Here they have been applied in two simple cases of one or two spacecraft experiencing indefinite fixed self-acceleration 
and a canonical launch-landing mission.
From these results, the query-response latencies are inferred directly through the composition of two appropriate
clock image functions.

The overall conclusion is that these round-trip latencies in both directions 
are a significant issue for interstellar travel,
and in fact such two-way communication is
for all practical purposes very cumbersome (due to large latencies) except in the vicinity of the \origin{} or the \destination{}.
Interstellar spacecraft and their crews must accept highly autonomous operations, and abandon notions of maintaining
operational and social interactions with those at the \origin{} or \destination{} throughout the mission, with the exception of a short period following launch or prior to landing.
Since the root cause is the large propagation distances, this general conclusion applies to many other 
types of interstellar missions largely irrespective of spacecraft speeds.

There are other issues in inter-traveler communications arising out of relativistic effects that have not been considered here.
These include aperture and antennae beam patterns as affected by 
a significant red shift in wavelength as well as relativistic aberration.
Gravitational effects, which have also been neglected, are a factor in the vicinity of \origin{} and \destination{}.

\section*{Acknowledgements}

The research represented in this tutorial review was supported by a grant from
the Breakthrough Foundation and its Breakthrough StarShot program,
NASA grants NIAC Phase I DEEP-IN – 2015 NNX15AL91G and 
NASA NIAC Phase II DEIS – 2016 NNX16AL32G and the 
NASA California Space Grant NASA NNX10AT93H, 
the Emmett and Gladys W. Technology Fund, and the Limitless Space Institute.

\appendix

\section{Nomenclature}
\label{sec:Nomenclature}

\begingroup \begin{center}
  \renewcommand*{\arraystretch}{1.3} \begin{longtable}{c p{10cm}}
    $S$ & A one-dimensional (rectilinear) inertial frame with map-position (spatial) and coordinate time \ile{\{x,t\}}
\\
$c$ & Speed of light in a vacuum as observed by any two synchronized coordinate-time clocks in an arbitrary inertial frame $S$
\\
\ile{\{\trav{A},\trav{B}\}} & Accelerated travelers observing traveler's times 
\ile{\{\travtime{a},\travtime{b}\}} (in a degenerate case, may also be at rest)
\\
$\trav{O}$ & Location of the origin of a spacecraft launch, assumed to be at rest in $S$
\\
$\trav{C}$ & A spacecraft, in motion and accelerating during its
cruise phase
\\
$\trav{D}$ & Location of the \destination{} for a spacecraft, assumed to be at rest in $S$
\\
$\distanceTarget$ & The fixed distance from $\trav{O}$ to $\trav{D}$
\\
MET & Mission elapsed time (initialized at the instant of launch)
\\
$\travtime{a}$ & Traveler's time measured by a clock carried by traveler $\trav{A}$
\\
METT & The MET as measured by the spacecraft traveler's clock
\\
$\coordtime{a}[\tau_a]$ & Coordinate time measured by a (typically notional) clock $\trav{A}$ at rest in an inertial frame $S$ at the instantaneous map-position of $\trav{A}$ at its traveler's time $\tau_a$
\\
MECT & The MET as measured by a (typically notional) coordinate clock at rest
at the instantaneous map-position of the spacecraft
\\
$\pm\alpha$ & A spacecraft traveler's self-acceleration (\ile{0{<}\alpha}); if positive, this is an acceleration, and if negative, this is
a deceleration
\\
$\tH$ & The time horizon of an accelerating spacecraft, equal to $\alpha/c$
\\
$g$ & Acceleration equal to that experienced at the Earth's surface due to gravity
\\
$\psidirection{A}{B}[\tau_a]$ & Clock image ($\trav{B}$'s traveler's arrival time) in photon exchange for a photon propagating in the direction $\direction{A}{B}$
that was emitted by $\trav{A}$ at its traveler's time $\tau_a$
\\
$\omegadirection{A}{B}[\tau_b]$ & Inverse clock image ($\trav{A}$'s traveler's photon emission time) in photon exchange for a photon propagating in the direction $\direction{A}{B}$
that arrives at $\trav{B}$'s
traveler's time $\tau_b$
\\
$\latency{A}{B}[\tau_a]$ & Query-response latency (elapsed traveler's time) for queries originating with $\trav{A}$ 
traveler's time $\tau_a$,
directed at $\trav{B}$ with an immediate response
back to $\trav{A}$
\\
$x_a[t_a[\tau_a]]$ & Map-position of $\trav{A}$ at coordinate time $\coordtime{a}[\tau_a]$ corresponding to traveler's
time $\tau_a$
\\
$r_a[\tau_a]$ & Map-position of $\trav{A}$ at traveler's time $\tau_a$, equal to
$x_a[t_a[\tau_a]]$
\\
$\gamma_a[\tau_a]$ & Time-speed (Lorentz factor) for $\trav{A}$ at traveler's
time $\tau_a$,
equal to the derivative of
$t_a[\tau_a]$ with respect to
$\tau_a$
\\
$w_a[\tau_a]$ & Map-speed of $\trav{A}$ as measured using map-position
$r_a[\tau_a]$ in conjunction with traveler's clock $\tau_a$
\\
$u_a[t[\tau]]$ & Map-speed of $\trav{A}$ as measured using map-position
$x_a[t_a[\tau_a]]$ in conjunction with coordinate clock $t_a[\tau_a]$, both measured
at traveler's time $\tau_a$
\\
$w_a^\prime[\tau_a]$ & Map-acceleration as measured using map--speed $w_a[\tau_a]$  in conjunction with traveler's clock $\tau_a$
\\
\ile{\{\trav{C1},\trav{C2}\}}
&
Two spacecraft launched at different times, with $\trav{C1}$ launched prior to $\trav{C2}$
\\
$\epsilon$ & For a second spacecraft $\trav{C2}$, the delay in its launch relative to
a first spacecraft $\trav{C1}$,
expressed as a fraction of the time horizon $\tH$ (so \ile{0{<}\epsilon{<}1})

   \end{longtable}
\end{center}
\endgroup

\section{Analytical results for indefinite acceleration}

The assumption of  indefinite fixed-magnitude self-acceleration yields straightforward
analytical results, which are now derived.

\subsection{Spacecraft trajectory}
\label{sec:trajConstAccel}

The trajectory of the spacecraft of \eqnref{motionVarsIndefiniteAccel} is now derived.
Solving differential equation \eqnref{selfToMapAccel} with initial condition \ile{w_c [0]{=}0}
and substituting the resulting $w_c [\travtime{c}]$ into \eqnref{Lorentz} yields
\begin{subequations}
\begin{align}
\label{eq:motionVarsMapSpeed}
&w_c[\travtime{c}] = c \cdot \sinh \big(\travtime{c}/\tH \big)
\\
\label{eq:motionVarsGamma}
&\gamma_c [\travtime{c}] = \cosh  \big(\travtime{c}/\tH \big)
\,,
\end{align}
\end{subequations}
Note that $c$ as a subscript in \eqnref{trajPhoton} is distinct from $c$ as a physical constant (the speed of light).
Solving the differential equation \ile{t^\prime [\travtime{c}] = \gamma_c [\travtime{c}]},
substituting \eqnref{motionVarsGamma} and
with initial condition \ile{\travtime{c} [0]{=}0},
yields \eqnref{motionTimeCoordTime}.

The map-position can be recast as
\begin{equation}
\label{eq:diffEqPosition}
r_c^\prime [\travtime{c}] = x_c^\prime \big[\travtime{c}[\travtime{c}]\big] \cdot \gamma_c [\travtime{c}] =  
u_c \big[\travtime{c} [\travtime{c}]\big] \cdot \gamma_c [\travtime{c}] 
= w_c [\travtime{c}] \,,
\end{equation}
from \eqnref{speedTravelers}.
The $r_c[\travtime{c}]$ in \eqnref{motionPositionCoordTime} is the solution to differential equation \eqnref{diffEqPosition} after
substituting from \eqnref{motionVarsMapSpeed}
with initial condition \ile{r_c[0]{=}0}.
The canonical mission trajectory is readily  inferred from this simpler mission (see \secref{trajectoryCanonical}).

\subsection{Clock images and query-response latency}
\label{sec:imageAnal}

Consider the trajectory of a photon emitted from  position \ile{\{0,t_0\}} in direction $\direction{O}{C}$ and intersecting the spacecraft trajectory at \ile{\{r_c[\travtime{c}],t_c[\travtime{c}]\}},
and also a photon emitted in direction $\direction{C}{O}$ from the spacecraft position  
\ile{\{r_c[\travtime{c}],t_c[\travtime{c}]\}}
that intersects \origin{} 
\ile{\{0,t_0\}}.
Then the respective photon trajectories, based on range equation \eqnref{range}, are
\begin{subequations}
\label{eq:trajPhoton}
\begin{align}
\label{eq:trajPhotonPlus}
&\direction{O}{C}\text{:  }
\coordtime{c}[\travtime{c}] =\coordtime{o} + r_c [\travtime{c}]/c
\\
\label{eq:trajPhotonMinus}
&\direction{C}{O}\text{:  }
\coordtime{o} = \coordtime{c}[\travtime{c}] + r_c [\travtime{c}]/c \,.
\end{align}
\end{subequations}

The clock images can be determined numerically
by calculating $\coordtime{o}$ for a set of different values of $\travtime{c}$ in \eqnref{trajPhoton} 
(and equivalent
relations for the \destination{}).
This yields a list of 
\ile{\{\travtime{o},\travtime{c},\travtime{d}\}} 
triplets which directly sample the relevant clock images.
This approach was used to generate \figref{clockImage1gVsDistanceUp}, \figref{clockImage1gVsDistanceDown},
\figref{responseImageUp} and \figref{responseImageDown}.

For simple trajectory \eqnref{motionVarsIndefiniteAccel}, analytical results are easily obtained,
and these provide additional insight. 
The clock images in \eqnref{imageConstAccel} are the solutions
(with the trajectory variables substituted from \eqnref{motionVarsIndefiniteAccel})
to \eqnref{trajPhotonPlus} for $\travtime{c}$
and to \eqnref{trajPhotonMinus} for \ile{\coordtime{o} \equiv \travtime{o}}.
Similarly, the inverse clock images in \eqnref{imageFns1} and \eqnref{imageFns2}
are the solutions to \eqnref{trajPhotonPlus} for  \ile{\coordtime{o} \equiv \travtime{o}}
and to \eqnref{trajPhotonMinus} for $\travtime{c}$.
The query-response latencies associated with the indefinite self-acceleration trajectory of \eqnref{latencyFixedAccel}
follow directly from \eqnref{imageConstAccel} substituted into \eqnref{defnlatency}.

\subsection{Communication between two accelerating spacecraft}
\label{sec:imageTwo}

Consider two spacecraft, each with indefinite acceleration as shown in
\figref{trajTwoRockets}, which permits communication between
the spacecraft in both directions under some conditions.
Analytical results for clock images are readily obtained as an extension of the range equation
technique used in \secref{imageAnal}.

The two spacecraft are labeled $\trav{C1}$ (the earliest launch) and $\trav{C2}$ (the later launch).
It is assumed that $\trav{C2}$ is launched at coordinate time \ile{\coordtime{o} = \epsilon \tH},
and as well that the
traveler's clock for $\trav{C2}$ is initialized to \ile{\travtime{c2}{=}0} at that instant of launch.
We must have \ile{0{<}\epsilon{<}1}, or else all photon emissions from $\trav{C2}$ violate
the event horizon of $\trav{C1}$ and thus will never be detected.
In that case the trajectory of $\trav{C2}$ is identical to that of $\trav{C1}$ except that it is shifted
up by $\epsilon \tH$ in the time direction.
Analogously to \eqnref{trajPhoton}, the three points at which
the trajectory of a photon emitted from $\trav{O}$
and travelling in the $+x$ direction intersects the trajectories of $\trav{C2}$ and $\trav{C1}$ are then given by
\begin{equation}
\coordtime{o}  = \coordtime{c1}[\travtime{c1}] - r_{c1} [\travtime{c1}]/c
= \epsilon \tH + \coordtime{c2}[\travtime{c2}] - r_{c2} [\travtime{c2}]/c \,.
\end{equation}
Substituting for \ile{\{ r_c, t_c\}} from \eqnref{motionTimeCoordTime} and
solving for each traveler's time in terms of the other yields the image function
for photon exchanges in the direction $\direction{C2}{C1}$,
\begin{subequations}
\label{eq:imageC2C1}
\begin{align}
\label{eq:imagePsiOC2}
&\psidirection{C2}{C1}[\travtime{c2}] = -\, \tH \cdot \log \left( \text e^{-\travtime{c2}/\tH} - \epsilon \right) 
\,,\ 0 \le \travtime{c2} < -\, \tH \cdot \log \epsilon
\\
\label{eq:imageOmegaOC2}
&\omegadirection{C2}{C1}[\travtime{c1}] =  -\, \tH \cdot \log \left( \text e^{-\travtime{c1}/\tH} + \epsilon \right)  
\,,\ 0 \le  \travtime{c1} \,.
\end{align}
\end{subequations}
There is an upper bound on the time over which $\trav{C2}$
can communicate with $\trav{C1}$; namely, only as long as $\trav{C2}$ remains within the time horizon of $\trav{C1}$.

Likewise, a photon emitted
in direction $\direction{C1}{C2}$ from either $\trav{C1}$
or $\trav{C2}$ and reaching $\trav{O}$ at the same $\coordtime{o}$
must satisfy the range equations
\begin{equation}
\coordtime{o} = \coordtime{c1}[\travtime{c1}] + r_{c1} [\travtime{c1}]/c
= \epsilon \tH + \coordtime{c2}[\travtime{c2}] + r_{c2} [\travtime{c2}]/c \,.
\end{equation}
Solving for the clock image functions in this direction,
\begin{subequations}
\label{eq:imageC1C2}
\begin{align}
\label{eq:imagePsiC2O}
&\psidirection{C1}{C2}[\travtime{c1}] =  \tH \cdot \log \left(  \text e^{\travtime{c1}/\tH} - \epsilon \right) 
\,,\ \tH \cdot \log (1 + \epsilon) \le  \travtime{c1}
\\
\label{eq:imageOmegaC2O}
&\omegadirection{C1}{C2}[\travtime{c2}] =  \tH \cdot \log \left(  \text e^{\travtime{c2}/\tH} + \epsilon \right) 
\,,\ 0 \le  \travtime{c2}
\,.
\end{align}
\end{subequations}
There is a lower limit on $\travtime{c1}$
since earlier emissions would arrive prior to launch.

The query-response latencies follow from substituting \eqnref{imageC2C1} and \eqnref{imageC1C2} into \eqnref{defnlatency},
\begin{subequations}
\begin{align}
\latency{C2}{C1}[\travtime{c2}] &=
\tH \cdot \log \left( \frac{1}{\text e^{-\travtime{c2}/\tH} - \epsilon} - \epsilon \right)
 - \travtime{c2}
 \,, \ 
0 \le \travtime{c2} < -\, \tH \cdot \log \epsilon
\\
\latency{C1}{C2}[\travtime{c1}] &= 
-\,\tH \cdot \log \left( \frac{1}{\text e^{\travtime{c1}/\tH} - \epsilon} - \epsilon \right)
 - \travtime{c1}
\,,
\\
\notag
&\ \ \ \ \ \ \ \ \ \ \ \ \ \ \ \ \ \ \ \ \ \ \ \ \ \ 
\tH \cdot \log (1 + \epsilon) \le  \travtime{c1} < \tH \cdot \log \left( \epsilon + \frac{1}{\epsilon} \right)
\,.
\end{align}
\end{subequations}
The upper bound on $\travtime{c1}$ in $\latency{C1}{C2}$ follows because the arrival time of a query at $\trav{C2}$
must fall within the allowed emission time given in \eqnref{imagePsiOC2}.
The upper bound as referred to $\travtime{c1}$ is found by applying the inverse clock image in \eqnref{imageOmegaC2O},
\begin{equation}
\omegadirection{C1}{C2} \big( -\, \tH \cdot \log \epsilon \big) = \tH \cdot \log \left( \epsilon + \frac{1}{\epsilon} \right)
\,.
\end{equation}
The resulting
$\latency{C2}{C1}$ and $\latency{C1}{C2}$ are plotted in \figref{twoRocketLatencyUp} and \figref{twoRocketLatencyDown}.

\subsection{Time warping derivatives}
\label{sec:timeWarp}

The derivatives in \eqnref{warpFactor} can be derived analytically by taking the total derivative of the
range equations \cite{RefNumber806}.
Alternatively they can be expressed in terms of clock images as
\begin{subequations}
\begin{align}
&\frac{\text d \travtime{c}}{\text d \travtime{o}} \bigg|_{\direction{O}{C}}
=  \bigg( \frac{\text d}{\text d \travtime{o}} \psidirection{O}{C} \big[\travtime{o}\big] 
\bigg)_{\travtime{0}\,=\,\omegadirection{O}{C} [\travtime{c}]}
\\
&\frac{\text d \travtime{o}}{\text d \travtime{c}} \bigg|_{\direction{C}{O}}
= \frac{\text d}{\text d \travtime{c}} \psidirection{C}{O} [\travtime{c}] 
\,.
\end{align}
\end{subequations}
Both derivatives are evaluated at the same
spacecraft traveler's time $\travtime{c}$.
Substituting from \eqnref{imageConstAccel}
and \eqnref{imageFns}, both derivatives evaluate to the same value,
that given in \eqnref{warpFactor}.
In addition, the ratio
of clock images in \eqnref{warpFactor}
evaluates to that same value
when substituted from
\eqnref{imageConstAccelCO} and \eqnref{imageFns1}.

\end{makefigurelist}

\end{document}